\pdfoutput=1
\documentclass[11pt,twoside,a4paper,cmspaper,final,collab]{cms-tdr}
\def\svnVersion{3fdb428}\def\svnDate{2025/10/01}\def\cmsCernNoTag{CERN-EP-2025-025}\def\cmsCernDate{\today}\def\cmsMessage{Published in the Journal of High Energy Physics as \href{http://dx.doi.org/10.1007/JHEP09(2025)141}{\doi{10.1007/JHEP09(2025)141}.}}
\begin{document}\cmsNoteHeader{SUS-23-004}

\newcommand{\PM}{\HepParticle{M}{}{}\xspace}
\newcommand{\PchiDM}{\HepParticle{\upchi}{}{}\xspace}
\newcommand{\PAchiDM}{\HepAntiParticle{\upchi}{}{}\xspace}
\newcommand{\massPM}{\ensuremath{m_{\PM}}\xspace}
\newcommand{\massPchiDM}{\ensuremath{m_{\PchiDM}}\xspace}
\newcommand{\pp}{\ensuremath{\Pp\Pp}\xspace}
\newcommand{\singlet}{\ensuremath{\text{single }\PQt}\xspace}
\newcommand{\tw}{\ensuremath{\PQt\PW}\xspace}
\newcommand{\ttbarlnu}{\ensuremath{\ttbar(\Pell\PGn)}\xspace}
\newcommand{\ttbarenu}{\ensuremath{\ttbar(\Pe\PGn)}\xspace}
\newcommand{\ttbarmunu}{\ensuremath{\ttbar(\PGm\PGn)}\xspace}
\newcommand{\vjets}{\ensuremath{\PV\text{+jets}}\xspace}
\newcommand{\wlnu}{\ensuremath{\PW(\Pell\PGn)}\xspace}
\newcommand{\wenu}{\ensuremath{\PW(\Pe\PGn)}\xspace}
\newcommand{\wmunu}{\ensuremath{\PW(\PGm\PGn)}\xspace}
\newcommand{\wjets}{\ensuremath{\PW\text{+jets}}\xspace}
\newcommand{\wlnujets}{\ensuremath{\wlnu\text{+jets}}\xspace}
\newcommand{\zll}{\ensuremath{\PZ(\Pell\Pell)}\xspace}
\newcommand{\znunu}{\ensuremath{\PZ(\PGn\PGn)}\xspace}
\newcommand{\zee}{\ensuremath{\PZ(\Pe\Pe)}\xspace}
\newcommand{\zmumu}{\ensuremath{\PZ(\PGm\PGm)}\xspace}
\newcommand{\zjets}{\ensuremath{\PZ\text{+jets}}\xspace}
\newcommand{\zlljets}{\ensuremath{\zll\text{+jets}}\xspace}
\newcommand{\znunujets}{\ensuremath{\znunu\text{+jets}}\xspace}
\newcommand{\gammajets}{\ensuremath{\PGg\text{+jets}}\xspace}
\newcommand{\ww}{\ensuremath{\PW\PW}\xspace}
\newcommand{\wz}{\ensuremath{\PW\PZ}\xspace}
\newcommand{\zz}{\ensuremath{\PZ\PZ}\xspace}
\newcommand{\abseta}{\ensuremath{\abs{\eta}}\xspace}
\newcommand{\mtw}{\ensuremath{m_{\mathrm{T},\PW}}\xspace}
\newcommand{\ptmissnomu}{\ensuremath{\pt^{\text{miss, no }\PGm}}\xspace}
\newcommand{\hadrecoil}{\ensuremath{R_{\mathrm{T}}}\xspace}
\newcommand{\hadrecoilvec}{\ensuremath{\vec{R}_{\mathrm{T}}}\xspace}
\newcommand{\mll}{\ensuremath{m_{\Pell\Pell}}\xspace}
\newcommand{\deltaeta}{\ensuremath{\Delta\eta}\xspace}
\newcommand{\deltaphi}{\ensuremath{\Delta\phi}\xspace}
\newcommand{\DeltaPhiEllPTMiss}{\ensuremath{\deltaphi(\Pell, \ptvecmiss)}\xspace}
\newcommand{\IRelMuon}{\ensuremath{I_{\text{rel}}^{\PGm}}\xspace}
\newcommand{\ptMuon}{\ensuremath{\pt^{\PGm}}\xspace}
\newcommand{\ptPhoton}{\ensuremath{\pt^{\PGg}}\xspace}
\newcommand{\ptPileup}{\ensuremath{\pt^{\text{pileup}}}\xspace}
\newcommand{\sumPtCharged}{\ensuremath{\sum\pt^{\text{charged}}}\xspace}
\newcommand{\sumPtNeutral}{\ensuremath{\sum\pt^{\text{neutral}}}\xspace}
\newcommand{\sumPtPhoton}{\ensuremath{\sum\ptPhoton}\xspace}
\newcommand{\uptype}{\ensuremath{U}\xspace}
\newcommand{\downtype}{\ensuremath{D}\xspace}
\newcommand{\gV}{\ensuremath{g_{\text{v}}}\xspace}
\newcommand{\gVU}{\ensuremath{\gV^{\uptype}}\xspace}
\newcommand{\gVD}{\ensuremath{\gV^{\downtype}}\xspace}
\newcommand{\gVUOneThree}{\ensuremath{(\gVU)_{13}}\xspace}
\newcommand{\gVchi}{\ensuremath{\gV^{\PchiDM}}\xspace}
\newcommand{\gA}{\ensuremath{g_{\text{a}}}\xspace}
\newcommand{\gAU}{\ensuremath{\gA^{\uptype}}\xspace}
\newcommand{\gAD}{\ensuremath{\gA^{\downtype}}\xspace}
\newcommand{\gAUOneThree}{\ensuremath{(\gVU)_{13}}\xspace}
\newcommand{\gAchi}{\ensuremath{\gA^{\PchiDM}}\xspace}
\newcommand{\gVorA}{\ensuremath{g_{\text{v}/\text{a}}}\xspace}
\newcommand{\gVorAU}{\ensuremath{\gVorA^{\uptype}}\xspace}
\newcommand{\gVorAD}{\ensuremath{\gVorA^{\downtype}}\xspace}
\newcommand{\sigmaB}{\ensuremath{\sigma\mathcal{B}}\xspace}
\newcommand{\relicdensity}{\ensuremath{\Omega_{\text{nbm}}h^{2}}\xspace}
\newcommand{\LikelihoodAI}{\ensuremath{\mathcal{L}_{A}}\xspace}
\newcommand{\nAIAlpha}{\ensuremath{n_{A}^{\alpha}}\xspace}
\newcommand{\nAIObs}{\ensuremath{n_{A}^{\text{obs}}}\xspace}
\newcommand{\LikelihoodBI}{\ensuremath{\mathcal{L}_{B}}\xspace}
\newcommand{\nBIBeta}{\ensuremath{n_{B}^{\beta}}\xspace}
\newcommand{\nBIObs}{\ensuremath{n_{B}^{\text{obs}}}\xspace}
\newcommand{\tBAIBetaAlpha}{\ensuremath{t}\xspace}
\newcommand{\rIAlpha}{\ensuremath{r^{\alpha}}\xspace}
\newcommand{\rZNuNu}{\ensuremath{r^{\znunu}}\xspace}
\newcommand{\rTTBar}{\ensuremath{r^{\ttbar}}\xspace}
\newcommand{\nuisparvec}{\ensuremath{\vec{\theta}}\xspace}
\newcommand{\pb}{\unit{pb}}
\newcommand{\valueLumi}{\ensuremath{138\fbinv}\xspace}
\newcommand{\valueSqrtS}{\ensuremath{13\TeV}\xspace}
\newcommand{\sqrtS}{\ensuremath{\sqrt{s}=\valueSqrtS}\xspace}
\newcommand{\progDeepJet}{\textsc{DeepJet}\xspace}
\newcommand{\progParticleNet}{\textsc{ParticleNet}\xspace}

\cmsNoteHeader{SUS-23-004}
\title{Search for dark matter production in association with a single top quark in proton-proton collisions at \texorpdfstring{\sqrtS}{sqrt(s) = 13 TeV}}

\author*[cms-collaboration]{The CMS Collaboration}
\date{\today}
\abstract{A search for the production of a single top quark in association with invisible particles is performed using proton-proton collision data collected with the CMS detector at the LHC at \sqrtS, corresponding to an integrated luminosity of \valueLumi. In this search, a flavor-changing neutral current produces a single top quark or antiquark and an invisible state nonresonantly. The invisible state consists of a hypothetical spin-1 particle acting as a new mediator and decaying to two spin-1/2 dark matter candidates. The analysis searches for events in which the top quark or antiquark decays hadronically. No significant excess of events compatible with that signature is observed. Exclusion limits at 95\% confidence level are placed on the masses of the spin-1 mediator and the dark matter candidates, and are compared to constraints from the dark matter relic density measurements. In a vector (axial-vector) coupling scenario, masses of the spin-1 mediator are excluded up to 1.85 (1.85)\TeV with an expectation of 2.0 (2.0)\TeV, whereas masses of the dark matter candidates are excluded up to 0.75 (0.55)\TeV with an expectation of 0.85 (0.65)\TeV.}

\hypersetup{%
pdfauthor={CMS Collaboration},
pdftitle={Search for dark matter production in association with a single top quark in proton-proton collisions at sqrt(s) = 13 TeV},
pdfsubject={CMS},
pdfkeywords={CMS, dark matter, mono-top}}

\maketitle

\section{Introduction}
\label{sec:introduction}

The standard model (SM) of particle physics is the most complete description of high-energy physics to date.
It describes phenomena with unprecedented accuracy, and it has been continuously validated by experimental measurements.
Nevertheless, multiple observations have been made that cannot be conclusively explained by the SM, such as invisible mass in the universe inferred from astrophysical~\cite{cite:rotation_curves_rubin_ford,cite:bullet_cluster} and cosmological~\cite{cite:Planck_cosmo_params} observations.
These observations are consistent with galaxies being predominantly composed of nonrelativistic, weakly interacting mass that has yet to be conclusively explained via beyond-the-SM (BSM) theories.
This matter, typically referred to as dark matter (DM), could be produced and observed in a laboratory environment.

A common approach to searching for DM at a collider experiment, such as CMS at the CERN LHC, is an approach known as the ``mono-\PX search strategy''.
This strategy is based on the assumed interactions of DM with SM particles and the conservation of momentum.
If DM is produced in the detector in association with an SM particle \PX, it creates missing transverse momentum (\ptmiss) representing an imbalance of total transverse momentum (\pt) in the event.
Many searches for DM~\cite{Beltran:2010ww,Fox:2011pm,Goodman:2010ku,Rajaraman:2011wf,ATLAS:2015qlt,Bai:2012xg,ATLAS:2013ndy,CMS:2014fjm,Bell:2012rg,Carpenter:2012rg,ATLAS:2014wfc,Carpenter:2013xra,Berlin:2014cfa,CMS:2014rwa,ATLAS:2014kci} rely on this quantity.
Depending on the particle \PX that is produced in association with the DM, different event signatures appear, \eg, DM in association with a highly energetic light-flavor quark or gluon jet~\cite{cite:monojet_CMS,cite:monojet_ATLAS}, a \PZ boson~\cite{CMS:2020ulv,ATLAS:2021gcn}, a Higgs boson~\cite{CMS:2018zjv,CMS:2019ykj}, a photon~\cite{CMS:2018ffd,ATLAS:2020uiq}, or a single top quark (mono-top)~\cite{cite:monotop_had_CMS_2016,ATLAS:2024xne}.

In this article, a search for DM production in association with a single top quark is presented.
The term top quark denotes both the top quark and the top antiquark, and DM production refers to the production of a pair of DM candidates in the \GeVns to \TeVns mass range via a mediator particle.
The analysis is based on the data set collected by the CMS experiment in proton-proton (\pp) collisions at a center-of-mass energy of \valueSqrtS from 2016 to 2018, corresponding to an integrated luminosity of \valueLumi.

The mono-top signature consists of a single top quark and a large amount of \ptmiss in the final state.
Mono-top production in the SM cannot occur at tree level, but only at one-loop level and at higher orders in quantum chromodynamics (QCD) perturbation theory.
Furthermore, Cabibbo suppression~\cite{cite:CKM_1,cite:CKM_2} as well as the GIM mechanism~\cite{cite:GIM} reduce the cross section for the production of this signature by SM processes.
Thus, searches for mono-top production in BSM theories benefit from low expected SM backgrounds.
This analysis focuses on the hadronic decay channel of the top quark, $\PQt \to \PW\PQb \to \PQq\PAQq^\prime\PQb$, and targets events in which the top quark obtains a significant recoil against the mediator.

\begin{figure}[!ht]
  \centering
  \includegraphics[width=0.49\textwidth]{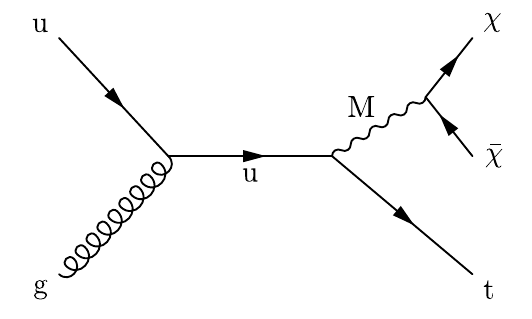}
  \caption{%
    Representative Feynman diagram of nonresonant mono-top production at tree level via a flavor-changing neutral current mediated by the spin-1 boson \PM.
    The off-shell up quark (\PQu) decays into an on-shell top quark (\PQt) and an \PM boson.
    The \PM boson decays directly to a pair of DM candidates \PchiDM and \PAchiDM.
  }
  \label{fig:monotop_feynman_diagram}
\end{figure}

The analysis is optimized for, and results of the search are interpreted in, a simplified model called the nonresonant mono-top model~\cite{cite:monotop_1}.
This simplified model describes the production of two spin-1/2 DM candidates (\PchiDM) in association with a top quark by introducing a flavor-changing neutral current (FCNC) process mediated by a new spin-1 boson (\PM).
By extending the SM with this model, nonresonant mono-top production becomes possible at tree level, as shown in Fig.~\ref{fig:monotop_feynman_diagram}.
Because the masses of the new particles are unknown, broad ranges of mediator and DM candidate masses are investigated, and purely vector, as well as purely axial-vector coupling scenarios, are tested.

The event selection is optimized for the mono-top signature.
A main event sample enriched in signal events is created by requiring a large-radius jet to capture the top quark, large recoil against the large-radius jet representing the spin-1 boson, and no additional activity in the event.
The signal purity is further enhanced by requiring the large-radius jet to be identified as originating from a top quark by a graph neural network~\cite{cite:particlenet}.
The most important backgrounds in this analysis are events from vector boson production in association with jets (\vjets) and top quark-antiquark pair (\ttbar) production.
Events from phase space regions not overlapping with the main sample are selected to create control samples enriched in these processes.
They are used to constrain the most important background contributions using data.

This article is structured as follows.
The CMS detector is described in Section~\ref{sec:cms_detector}, and the event and object reconstruction methods are explained in Section~\ref{sec:event_reconstruction}.
The signal model is introduced in Section~\ref{sec:signal_model}. The data as well as the signal and background simulations are described in Section~\ref{sec:datasamples}.
The event selection is described in Section~\ref{sec:event_selection} and the methodology of the top quark tagging is covered in Section~\ref{sec:top_quark_tagging}.
The signal extraction and the background estimation techniques are given in Section~\ref{sec:background_estimation}.
The systematic uncertainties are detailed in Section~\ref{sec:systematic_uncertainties}.
The analysis results and interpretations are presented in Section~\ref{sec:resultsandinterpretation}, and the article is summarized in Section~\ref{sec:summary}.
Tabulated results of this analysis are provided in the HEPData record~\cite{HEPData}.

\section{The CMS detector}
\label{sec:cms_detector}

The central feature of the CMS apparatus is a superconducting solenoid of 6\unit{m} internal diameter, providing a magnetic field of 3.8\unit{T}.
Within the solenoid volume are a silicon pixel and strip tracker, a lead tungstate crystal electromagnetic calorimeter (ECAL), and a brass and scintillator hadron calorimeter (HCAL), each composed of a barrel and two endcap sections.
Forward calorimeters extend the pseudorapidity ($\eta$) coverage provided by the barrel and endcap detectors.
Muons are measured in gas-ionization detectors embedded in the steel flux-return yoke outside the solenoid.
Events of interest are selected using a two-tiered trigger system.
The first level (L1), composed of custom hardware processors, uses information from the calorimeters and muon detectors to select events at a rate of around 100\unit{kHz} within a fixed latency of about 4\mus~\cite{CMS:2020cmk}.
The second level, known as the high-level trigger, consists of a farm of processors running a version of the full event reconstruction software optimized for fast processing and reduces the event rate to around 1\unit{kHz} before data storage~\cite{CMS:2016ngn, CMS:2024aqx}.
The primary vertex is taken to be the vertex corresponding to the hardest scattering in the event, evaluated using tracking information alone, as described in Section 9.4.1 of Ref.~\cite{CMS-TDR-15-02}.
A more detailed description of the CMS detector, together with a definition of the coordinate system used and the relevant kinematic variables, can be found in Refs.~\cite{CMS:2008xjf,CMS:2023gfb}.

\section{Event reconstruction}
\label{sec:event_reconstruction}

A particle-flow algorithm~\cite{CMS:2017yfk} aims to reconstruct and identify each individual particle in an event, with an optimized combination of information from the various elements of the CMS detector.
The energy of photons is obtained from the ECAL measurement.
The energy of electrons is determined from a combination of the electron momentum at the primary vertex, as determined by the tracker, the energy of the corresponding ECAL cluster, and the energy sum of all bremsstrahlung photons spatially compatible with originating from the electron track.
The momentum of muons is measured in the inner tracker and the muon chambers.
The energy of charged hadrons is determined from a combination of their momentum measured in the tracker and the matching ECAL and HCAL energy deposits, corrected for the response function of the calorimeters to hadronic showers.
Finally, the energy of neutral hadrons is obtained from the corresponding corrected ECAL and HCAL energies.

For each event, hadronic jets are clustered from these reconstructed particles using the infrared- and collinear-safe anti-\kt algorithm~\cite{Cacciari:2008gp, Cacciari:2011ma} with a distance parameter of 0.4 or 1.5, referred to as AK4 and AK15 jets, respectively.
Jet momentum is determined as the vectorial sum of all particle momenta in the jet, and is found from simulation to be, on average, within 5 to 10\% of the true momentum over the entire range and detector acceptance.
Additional \pp interactions within the same or nearby bunch crossings (pileup) can contribute additional tracks and calorimetric energy depositions to the jet momentum.
To mitigate this effect for AK4 jets, charged particles identified to be originating from pileup vertices are discarded and an offset correction is applied to adjust for remaining contributions~\cite{CMS:2020ebo}.
AK15 jets have a larger area and are more sensitive to the effects of pileup.
For this reason, the pileup-per-particle identification algorithm~\cite{CMS:2020ebo,Bertolini:2014bba} is used to mitigate the effect of pileup for AK15 jets at the reconstructed-particle level, making use of local shape information, event pileup properties, and tracking information.

Jet energy corrections are derived from simulation studies so that the average measured energy of jets becomes identical to that of generated-particle level jets.
In situ measurements of the momentum balance in \gammajets, \zjets events, as well as events composed uniquely of jets produced through the strong interaction (QCD multijet events), are used to account for any residual differences in the jet energy scale between data and simulation~\cite{CMS:2016lmd}.
For jets with central \abseta, the jet energy resolution ranges from around 15\% at 30\GeV, to 10\% at 100\GeV, and 5\% at 1\TeV~\cite{CMS:2016lmd}.
Additional selection criteria are applied to each jet to remove jets that are potentially dominated by anomalous contributions from various sub-detector components or reconstruction failures~\cite{CMS:2019ctu}.

For electrons, photons, and muons, a loose and a tight selection is established.
Loose objects are selected with a high identification efficiency and are used to veto events with additional objects in the final state.
Tight objects are selected with a lower misidentification rate compared to the loose selection.
They are used to select dedicated event samples, enriched with events from electroweak (EW) processes and \ttbar production with prompt electrons, muons, and photons in the final state.
The tight collections are always subsets of the corresponding loose collections.

Tight (loose) electrons must fulfill $\pt > 40$ (10)\GeV and $\abseta < 2.4$ (2.5).
A sequence of additional selection criteria is applied to improve the correct identification of genuine electrons, with an efficiency of around 70 (95)\% for tight (loose) electrons~\cite{CMS:2020uim}.
Tight (loose) photons must fulfill $\pt > 200$ (20)\GeV and $\abseta < 1.479$ (2.5).
In the 2016 data-taking period, the \pt threshold for selecting tight photons is increased to $\pt > 230\GeV$.
Additional requirements are imposed in order to improve the identification of genuine photons, which lead to an identification efficiency of 80 (90)\% for tight (loose) photons~\cite{CMS:2020uim}.

Tight (loose) muons must fulfill $\pt > 30$ (20)\GeV and $\abseta < 2.4$.
In the 2018 data-taking period, the \pt threshold for selecting loose muons is lowered to $\pt > 15\GeV$, while all other selection criteria stay the same.
The isolation variable
\begin{equation}
  \IRelMuon = \frac{1}{\ptMuon}\left[ \sumPtCharged + \max\left( 0, \sumPtNeutral + \sumPtPhoton - \ptPileup \right) \right]
\end{equation}
is employed to increase the selection purity of prompt muons produced in the hard process.

The scalar \pt sums of charged particles, neutral hadrons, and photons, denoted as \sumPtCharged, \sumPtNeutral, and \sumPtPhoton, respectively, are calculated for reconstructed particle candidates in a cone of radius $\DR = \sqrt{\smash[b]{(\deltaeta)^2 + (\deltaphi)^2}} < 0.4$ around the muon.
Here, \deltaeta and \deltaphi correspond to the difference in $\eta$ and in the azimuthal angle $\phi$ (measured in radians) between the muon and a reconstructed particle candidate.
The muon transverse momentum \ptMuon is excluded from these calculations.
The estimated contribution from pileup (\ptPileup) is subtracted from $\sumPtNeutral + \ptPhoton$.
Tight (loose) muons must fulfill $\IRelMuon < 0.15$ (0.25).
The muon identification has an efficiency of 96.0 (99.8)\% for tight (loose) muons~\cite{CMS:2023dsu}.

The most important selection criteria for electrons, photons, and muons in this analysis are summarized in Table~\ref{tab:reconstruction:electron_muon_photon_selection}.

\begin{table}[!ht]
  \centering
  \topcaption{%
    Requirements on electrons, photons, and muons that pass the loose or tight selection.
    For all objects, the minimal \pt, the maximal \abseta, and the efficiency of the object identification (ID) are provided.
    For muons, the requirements on the relative isolation \IRelMuon are also listed.
    A more detailed discussion is given in the text.
  }%
  \label{tab:reconstruction:electron_muon_photon_selection}%
  \renewcommand{\arraystretch}{1.5}
  \begin{tabular}{lcccc}%
                                               & \pt greater than     & \abseta less than      & ID efficiency            & \IRelMuon less than        \\
    \hline
    Loose electrons                            & 10\GeV               & 2.5                    & 95\%                     & \NA                        \\
    Tight electrons                            & 40\GeV               & 2.4                    & 70\%                     & \NA                        \\
    Loose photons                              & 20\GeV               & 2.5                    & 90\%                     & \NA                        \\
    \multirow{2}{*}{Tight photons}             & 230\GeV (2016),      & \multirow{2}{*}{1.479} & \multirow{2}{*}{80\%}    & \multirow{2}{*}{\NA}       \\[-1.0ex]
                                               & 200\GeV (2017, 2018) &                          &                            &                            \\
    Loose muons                                & 10\GeV               & 2.4                    & 99.8\%                  & 0.25                     \\
    \multirow{2}{*}{Tight muons}               & 20\GeV (2016, 2017), & \multirow{2}{*}{2.4}   & \multirow{2}{*}{96.0\%} & \multirow{2}{*}{0.15}    \\[-1.0ex]
                                               & 15\GeV (2018)        &                          &                            &                            \\
  \end{tabular}%
\end{table}

The missing transverse momentum vector \ptvecmiss is computed as the negative vector sum of the transverse momenta of all the particle-flow candidates, in an event, including electrons, muons, photons, and charged and neutral hadrons.
Its magnitude is denoted as \ptmiss~\cite{CMS:2019ctu}.
The \ptvecmiss is modified to account for corrections to the energy scale of the reconstructed jets in the event.

The hadronic recoil \hadrecoilvec represents the recoil against the hadronic activity in the transverse plane of the event.
In this analysis, the hadronic recoil is defined as
\begin{equation}
    \label{eqn:event_reconstruction:hadrecoil}
    \hadrecoilvec = \ptvecmiss + \sum_{i}\vec{p}_{\text{T},i},
\end{equation}
where the sum runs over all electrons, muons and photons in the event, which pass the loose selections.
Its magnitude is denoted as \hadrecoil.

The quantity \ptmissnomu is used in the trigger and is computed using the same method as for \ptmiss, but excluding muon candidates in the trigger reconstruction from the calculation.
If no other electrons or photons are present in the final state, \ptmissnomu is very similar to the hadronic recoil.
A trigger that selects events based on \ptmissnomu directly targets the phase space of the search in the signal region (SR) and some of the control regions (CRs) defined in Section~\ref{sec:event_selection}.

Light flavor quarks (\PQu, \PQd, \PQs) and gluons produced in the primary vertex fragment and deposit clusters of energy in the ECAL and HCAL.
However, bottom quarks (\PQb) form semi-stable mesons that travel a finite distance before they decay.
The \progDeepJet tagger~\cite{cite:deepjet} is used to select jets that are consistent with a \PQb quark at a working point corresponding to a 10\% mistagging probability of light-flavor jets.
The efficiency to identify jets originating from a \PQb quark is approximately 90\% at this working point.
The selection efficiency and mistag rate of the \progDeepJet tagger in simulation are corrected with scale factors measured in data to account for residual differences between data and simulation~\cite{cite:deepjet}.

In the signal topology, the top quark (\PQt) recoils off the DM candidates in the event.
This causes the top quark to have a significant boost that allows for its decay products to be clustered into a single large-radius jet.
A graph neural network approach (\progParticleNet)~\cite{cite:particlenet} is used to assign probability-like scores to the possible origins (top quark, Higgs boson, EW bosons, QCD multijet) of large-radius jets.
The output of \progParticleNet is used to select events in which the large-radius jet likely originated from a hadronically decaying top quark.

\section{Signal model}
\label{sec:signal_model}

The analysis searches for a single top quark produced via an FCNC process.
The top quark is produced in association with a spin-1 boson that has FCNC couplings to quarks and can decay directly into DM candidates.
Figure~\ref{fig:monotop_feynman_diagram} depicts a representative Feynman diagram of this process.
This production channel is also called nonresonant mono-top production.

The effective Lagrangian allowing nonresonant mono-top production at tree level in the simplified model~\cite{cite:monotop_1} studied in this paper is given by
\begin{equation}
  \begin{aligned}
    \mathcal{L}
    =
      \mathcal{L}_{\text{SM}}
      & + \underbrace{\PM_\mu\overline{\uptype}\big(\gVU\gamma^\mu+\gAU\gamma^\mu\gamma_5\big)\uptype}_{\text{FCNC up-type \eg\ } \PQu \to \PQt}
        + \underbrace{\PM_\mu\overline{\downtype}\big(\gVD\gamma^\mu + \gAD\gamma^\mu\gamma_5\big)\downtype}_{\text{FCNC down-type \eg\ }\PQd \to \PQb} \\
      & + \underbrace{\PM_\mu\PAchiDM\big(\gVchi\gamma^\mu + \gAchi\gamma^\mu\gamma_5\big)\PchiDM}_{\text{decay }\PM \to \PchiDM\PAchiDM} \\
      & + \mathcal{L}_\text{kin, mass, gauge}\big( \PM_\mu, \massPM, \PchiDM, \massPchiDM \big).
  \end{aligned}
  \label{eq:monotop_lagrangian}
\end{equation}
The SM Lagrangian $\mathcal{L}_{\text{SM}}$ is extended by the mass, kinetic, and gauge terms of the newly introduced fields, as well as a term describing the interactions between the spin-1 mediator \PM with mass \massPM and SM quarks or the DM candidates \PchiDM with mass \massPchiDM.
The \PchiDM is assumed to be stable and the width of the mediator is dependent on the \massPM, \massPchiDM, and the axial and axial-vector couplings as defined in~\cite{Boveia:2016mrp}.
Here, \uptype represents the up-type quarks $(\PQu, \PQc, \PQt)^T$ and \downtype the down-type quarks $(\PQd, \PQs, \PQb)^T$ with the color indices implied, \gVchi (\gAchi) is the vector (axial-vector) coupling to DM candidates, and \gVU, \gAU, \gVD, \gAD are called flavor matrices, parameterizing the interaction strength of the different up/down-type quarks with the \PM boson.
The subscripts ``v'' and ``a'' represent the vector and axial-vector couplings.
The diagonal and nondiagonal elements of the flavor matrices represent the interaction strengths between up-type and down-type quarks of the same and different flavor, respectively.
Depending on which elements of the flavor matrices are chosen as nonzero, different mono-\PX signatures can appear.

The mono-top model described here is specified such that an FCNC connects the up-type quarks of the first and third generations, allowing mono-top production at tree level.
This implies that only $(\gVorAU)_{13} \neq 0$, whereas all other coupling elements of the up-type flavor matrices are set to zero.
This is motivated by the fact that top quark production through an $s$-channel up quark is favored by parton distribution functions (PDFs) at the LHC.
For SM gauge invariance, the same-generation down-type couplings need to be similar to the up-type couplings.
This means that also $(\gVorAD)_{13} \neq 0$, whereas all other coupling elements of the down-type flavor matrices are set to zero.

For the nominal coupling scenario of this analysis, \ie, a vector-only coupling scenario, all axial-vector couplings are set to zero, and $\gVUOneThree = 0.25$ and $\gVchi = 1$ is chosen, following the recommendations of the LHC Dark Matter working group~\cite{cite:lhcdmwg}.
Larger values of \gVUOneThree enhance decays of the mediator into bottom or down quarks, which are also kinematically favored.
This would suppress the branching fraction of the mediator's decay into DM candidates.
An alternative axial-vector coupling scenario, for which all vector couplings are set to zero and $\gAUOneThree = 0.25$ and $\gAchi = 1$ is chosen, is also considered.

\section{Data and simulated samples}
\label{sec:datasamples}

The data set used in the analysis was collected with the CMS experiment in \pp collisions in 2016, 2017, and 2018 and corresponds to a total integrated luminosity of \valueLumi at a center-of-mass energy of \valueSqrtS.

Signal and background events are simulated at leading order (LO) with \PYTHIA (v.~8.240)~\cite{Sjostrand:2014zea}, or at next-to-LO (NLO) in perturbative QCD with \MGvATNLO (v.~2.6.5)~\cite{Alwall:2014hca} or \POWHEG (v.~2)~\cite{powheg:general:1,powheg:general:2,powheg:general:3}, depending on the process, as defined below.
The NNPDF (v.~3.1) PDFs~\cite{NNPDF:2017mvq} at next-to-NLO (NNLO) are used for all processes.
All generated events are interfaced with \PYTHIA (v.~8.240) with the CP5 tune~\cite{CMS:2019csb} to generate the parton shower and to model the hadronization.
A detailed simulation of the response of the CMS detector is performed with \GEANTfour~\cite{GEANT4:2002zbu} for each event.
For all simulations, the value of the top quark mass is set to 172.5\GeV.

The signal is simulated at NLO QCD perturbation theory using \MGvATNLO~\cite{Alwall:2014hca} based on the nominal coupling scenario described in Section~\ref{sec:signal_model}.
The implementation of the model~\cite{cite:DMsimp} is provided by the authors of Ref.~\cite{cite:monotop_1}.
Alternative coupling scenarios are realized via a reweighting procedure in \MGvATNLO~\cite{Mattelaer:2016gcx}.
Signal samples in the range of mediator mass $200 \leq \massPM \leq 2500\GeV$ and DM candidate mass $50 \leq \massPchiDM \leq 1500\GeV$ have been produced to account for the unknown masses of the newly introduced particles.
Only on-shell decays of the mediator into DM candidates are considered, \ie, the relation $2\massPchiDM \leq \massPM$ must be fulfilled for a mass hypothesis to be taken into account.
The lower threshold on the mediator mass is chosen such that the mediator mass is always greater than the mass of the top quark to avoid new decay channels of the top quark mediated by the \PM boson.

The \ttbar background process is simulated using \POWHEG at NLO accuracy in the five-flavor scheme~\cite{Frixione:2007nw}.
The total cross section is normalized to calculations at NNLO precision in QCD including soft gluon resummation of next-to-next-to-leading logarithmic terms, obtained with \textsc{top++} (v.~2.0)~\cite{bib:ttbarxs:1,bib:ttbarxs:2,bib:ttbarxs:3,bib:ttbarxs:4,bib:ttbarxs:5,bib:ttbarxs:6,bib:ttbarxs:7}.

Backgrounds from single top quark (\singlet) production in the $t$ channel, and in association with a \PW boson (\tw), are simulated with \POWHEG~\cite{powheg:singlet:st,powheg:singlet:tW}.
Single \PQt production in the $s$ channel is simulated with \MGvATNLO.
The inclusive production cross sections are normalized to NLO calculations for \singlet production in the $t$ and $s$ channels~\cite{bib:singletxs:t:1,bib:singletxs:t:2}, and to approximate NNLO calculations for \tw production~\cite{bib:singletxs:tw:1}.

Further background contributions arise from \vjets production where the boson \PV represents a \PW, \PZ, or \PGg boson.
They are simulated with \MGvATNLO at NLO QCD precision with up to two additional partons in the final state, where the matching of matrix-element jets with those from parton showers is performed using the FxFx~\cite{Frederix:2012ps} prescription.
The \gammajets process is generated at NLO accuracy in perturbative QCD with up to one parton in the matrix element calculation.
A reweighting procedure based on the vector boson \pt is used in order to apply EW corrections at NLO precision, following Ref.~\cite{Lindert:2017olm}.
This improves the SM background prediction in phase space regions where contributions from \vjets production are dominant.
These EW corrections become relevant when the vector boson has a large transverse momentum.

Additional minor background contributions arise from diboson (\ww, \wz, and \zz) as well as QCD multijet production.
These are simulated at LO using the \PYTHIA event generator. The \ww and \zz processes are normalized to the NNLO cross sections~\cite{Gehrmann:2014fva,Cascioli:2014yka} and the \wz process is scaled to the NLO calculation~\cite{Campbell:2011bn}.
QCD multijet production is scaled to the LO cross section extracted from the generated event sample.

\section{Event selection}
\label{sec:event_selection}

The selection targets events with a significant amount of \hadrecoil.
The selected events are subsequently divided into SRs with a high selection efficiency for the signal, and CRs, which are depleted of signal and enriched in background processes.
The CRs target processes equal or similar to the most important backgrounds in the SR.
They take an important role in the background estimation of the \znunujets and \ttbar processes, which is further explained in Section~\ref{sec:background_estimation}.
To estimate the \znunujets shape in the SR, the method exploits the similarity of this process to \zlljets, \wlnujets, and \gammajets production.

Three triggers are used to select events in this analysis.
A trigger that is activated on events with a large amount of \ptmissnomu is used to collect events in the SRs and in CRs of background processes with muons in the final state.
In order to collect events in the CRs of background processes with electrons (photons) in the final state, an electron (a photon) trigger is used.

The signal events contain a hadronically decaying top quark and large \ptmiss.
As a phase space with top quarks with a high Lorentz boost is targeted, the decay products of the top quark can be reconstructed in a single AK15 jet.
Events considered in this analysis must fulfill $\hadrecoil > 350\GeV$.
This selection on \hadrecoil ensures that the analysis selects events in which the \ptmissnomu trigger efficiency is significantly greater than 95\% and removes events with low \hadrecoil that typically have a worse signal-to-background ratio.
Events must have at least one large-radius AK15 jet with $\pt > 250\GeV$ and $\abseta < 2.4$.
The leading AK15 jet, \ie, the AK15 jet with the highest \pt in the event, and the \hadrecoilvec must fulfill $\abs{\deltaphi} > 1.5$.
The AK4 jets that are considered in this analysis are required to have $\pt > 30\GeV$, $\abseta < 2.4$, and to not originate from detector noise.
They are mainly used to identify \PQb jets outside the leading AK15 jet in the event.
For events to be further considered, all AK4 jets, including those which overlap with the leading AK15 jet, must have a \deltaphi separation from the \hadrecoilvec by at least 0.5 radians.
This requirement heavily reduces contributions from QCD multijet production in which the \ptmiss is typically mismeasured collinearly with one of the leading AK4 jets.

In a large fraction of the 2018 data-taking period, two HCAL endcap modules (HEM) were not functioning, thereby impacting the jet energy measurements in the affected detector regions.
In order to account for this HEM effect, events are vetoed if any AK4 jet fulfilling a very loose set of selections is found in the corresponding detector regions $-1.57 < \phi < -0.87$ and $-3.0 < \eta < -1.3$.
To further reduce events in which a jet is fully lost due to the HEM issue, contributing at low values of \ptmiss, events are vetoed if $\ptmiss < 470\GeV$ and $-1.62 < \phi(\ptvecmiss) < -0.62$, following Ref.~\cite{cite:monojet_CMS}.

In this analysis, only AK4 jets that are well separated from the leading AK15 jet in the event are considered for \PQb tagging.
The purpose of these \PQb-tagged jets is to either veto or explicitly allow for additional \PQb quarks in the event, which do not originate from the decay of the boosted top quark.
To consider an AK4 jet as \PQb tagged in this analysis, it must have a \DR distance of more than 1.5 from the leading AK15 jet and it must fulfill the requirements of the loose working point of the \progDeepJet tagger~\cite{cite:deepjet} corresponding to a mistagging probability of light-flavor quark and gluon jets of approximately 10\%.

Events that fulfill the preselection and have passed the \ptmissnomu, the electron, or the photon triggers are further categorized.
An SR is formed in order to create an event sample with a high selection efficiency on signal events and a high rejection rate of events from reducible background processes, \eg, \wlnujets and \ttbar production.
In addition, different CRs are created, which are enriched in events from specific background processes.
The different CRs target \zlljets, \wlnujets, \gammajets, and \ttbarlnu production.
The SR and CRs are mutually exclusive, \ie, each selected event is uniquely assigned to a single analysis region.

For the SR, a data set collected with the \ptmissnomu trigger is used.
Events in the SR must not contain any electrons, muons, or photons from the loose collections, as well as any \PQb-tagged jet.

The CRs enriched in events from \zlljets, \wlnujets, and \ttbar production are formed by selecting events with leptons in the final state.
For each of these leptonic CRs, two versions exist: one with electrons and one with muons in the final state.
In the CRs with electrons in the final state, events must have activated the single electron trigger, while in CRs with muons in the final state, events must have activated the \ptmissnomu trigger.
In these CRs, \ptmissnomu corresponds to the definition of \hadrecoil in Section~\ref{sec:event_reconstruction}, which motivates the use of the \ptmissnomu trigger.
In addition, a CR enriched in prompt \PGg production is formed by selecting events with a photon in the final state from the photon-triggered data set.

Events in the \zll CR require exactly two same-flavor leptons with an invariant mass of the two-lepton system \mll near that of the nominal \PZ boson mass~\cite{ParticleDataGroup:2024cfk}, \ie, $60 < \mll < 120\GeV$, and low \ptmiss.
The requirements on well-reconstructed electrons and muons are slightly relaxed in the \zll CR, \ie, the loose electron and muon collections are used for event selection, to mitigate the low branching fractions of the \PZ boson decay into charged leptons.
In the \wlnu CR, a single tight lepton, significant \ptmiss, and no additional \Pb-tagged jets in the event are required.
Except for the requirement of at least one additional \PQb-tagged jet in the event, the selection criteria of the \ttbarlnu CR are the same as those of the \wlnu CR.
Events in the \PGg CR require one isolated photon and no additional \Pb-tagged jets.
If an event passes all selections, but contains additional loose electrons, muons, or photons compared to the aforementioned selections, it is removed from the analysis data set.

In the \ttbarlnu and \wlnu CRs, an additional requirement on the transverse mass of the \PW boson, defined as
\begin{equation}
  \mtw = \sqrt{2p_{\mathrm{T},\Pell}\ptmiss\big[1 - \cos\DeltaPhiEllPTMiss \big]},
\end{equation}
is introduced to ensure an orthogonal selection to a potential mono-top search in the leptonic top quark decay channel with the same data set.
Here, \DeltaPhiEllPTMiss is the azimuthal difference between the selected tight lepton (electron or muon) \Pell and \ptvecmiss.
Since mono-top signatures with a leptonic top quark decay lead to values of \mtw well above the \PW boson mass, events in the \ttbarlnu and \wlnu CRs are required to fulfill $\mtw < 150\GeV$.

A summary of the selections can be found in Table~\ref{tab:hadronic_selection}.

\begin{table}[!ht]
  \centering
  \topcaption{%
    Overview of the selections for the SR and the CRs, including the preselection.
    In the \zll CRs, loose leptons are used for the selection of additional objects in the final state.
    In all other CRs, leptons and photons from the tight collections are used to determine the number of additional objects. CRs with final state leptons are defined separately for electrons and muons.
    The first four selections are common to the SR and all CRs.
  }%
  \bgroup%
  \renewcommand{\arraystretch}{1.25}
  \begin{tabular}{cccccc}%
    Observable                                          & SR & \zll                         & \wlnu            & \ttbarlnu         & \PGg   \\
    \hline
    \hadrecoil                                          & \multicolumn{5}{c}{$>$350\GeV}                                                        \\
    Leading AK15 jet \pt                                & \multicolumn{5}{c}{$>$250\GeV}                                                        \\
    $\deltaphi(\text{AK15 jet}, \hadrecoilvec)$         & \multicolumn{5}{c}{$>$1.5}                                                            \\
    $\min\deltaphi(\text{AK4 jet}, \hadrecoilvec)$      & \multicolumn{5}{c}{$>$0.5}                                                            \\
    Additional objects                                  & \NA & 2 \Pell (both loose)            & 1 \Pell         & 1 \Pell            & 1 \PGg \\
    \PQb-tagged AK4 jets                                & 0   & \NA                             & 0               & $\geq$1            & 0      \\
    \ptmiss                                             & \NA & $<$120\GeV                        & $>$150\GeV      & $>$150\GeV         & \NA    \\
    \mtw                                                & \NA & \NA                             & $<$150\GeV      & $<$150\GeV         & \NA    \\
    \mll                                                & \NA & $60 < \mll < 120\GeV$           & \NA             & \NA                & \NA    \\
  \end{tabular}%
  \egroup
  \label{tab:hadronic_selection}%
\end{table}

\begin{figure}[!ht]
  \centering
  \includegraphics[width=0.49\textwidth]{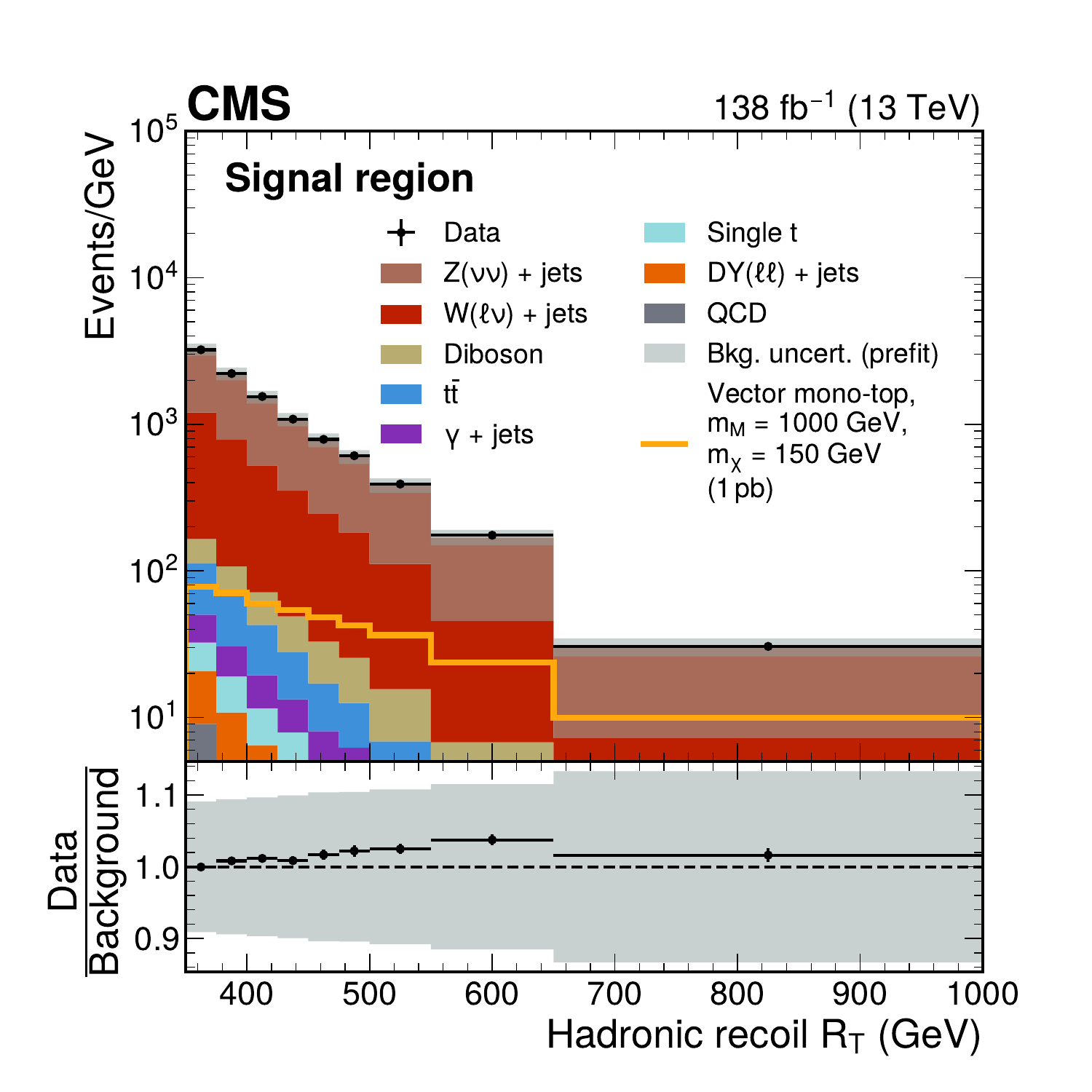}
  \caption{%
    Prefit distribution of the magnitude of the hadronic recoil \hadrecoil in the SR.
    The last bin of the distribution also contains events with $\hadrecoil > 1000\GeV$.
    The distributions of background processes from simulation are stacked together.
    All nominal corrections are applied to the simulated SM processes.
    A representative mono-top signal (vector coupling scenario) with a mediator mass of 1\TeV, a DM candidate mass of 150\GeV, and a cross section of 1\pb is overlaid as an orange line.
    The gray band represents the statistical and unconstrained systematic uncertainties in the simulated event yields.
  }%
  \label{fig:prefit_recoil}
\end{figure}

The distribution of the magnitude of \hadrecoil in the SR before the simultaneous binned maximum likelihood fit (prefit) is given in Fig.~\ref{fig:prefit_recoil}.
All background distributions are extracted from simulation, and all corrections and uncertainties, mentioned in Section~\ref{sec:systematic_uncertainties}, are applied.
As \hadrecoil represents the recoil against the top quark in the event, it is a measure for the mediator \pt in the signal process.
It is clearly visible that the mono-top signal tends to have larger \hadrecoil values than the expected background.
The most important background processes are \znunujets and \wlnujets production, in which the lepton from the \PW boson decay is either out of acceptance, not reconstructed, or not identified.
Both the SR and CRs will use top quark tagging, as detailed in Section~\ref{sec:top_quark_tagging}, to increase the sensitivity to the signal process and to improve the modeling of important SM backgrounds.

\section{Top quark tagging}
\label{sec:top_quark_tagging}

For the \vjets background processes, the large-radius jet originates from QCD radiation processes, while for the signal process it is mostly produced in the decay of the top quark. This property is exploited to reduce the contributions of the \vjets background processes to the SR.

For AK15 jets, a discriminator (top quark tagger, \PQt tagger), based on the graph neural network \progParticleNet, is employed to distinguish jets originating from the hadronic decay of a top quark from jets due to QCD radiation.
This discriminator provides probability-like scores $p_P$ for an AK15 jet to stem from a certain physics process $P$, \eg, from a hadronic top quark decay ($p_{\PQt}$), from pure QCD radiation ($p_{\text{QCD}}$), or from other possible origins that are not relevant for this analysis.
As discriminant, the aggregated score
\begin{equation}
  \label{eq:top_quark_tagging:t_vs_qcd}
  p_{\PQt/\text{QCD}} = \frac{p_{\PQt}}{p_{\PQt} + p_{\text{QCD}}}
\end{equation}
is used, which expresses the probability that the AK15 jet originates from a hadronic top quark decay as opposed to the hypothesis that the jet is purely QCD-initiated.

\begin{figure}[!ht]
  \centering
  \includegraphics[width=0.7\textwidth]{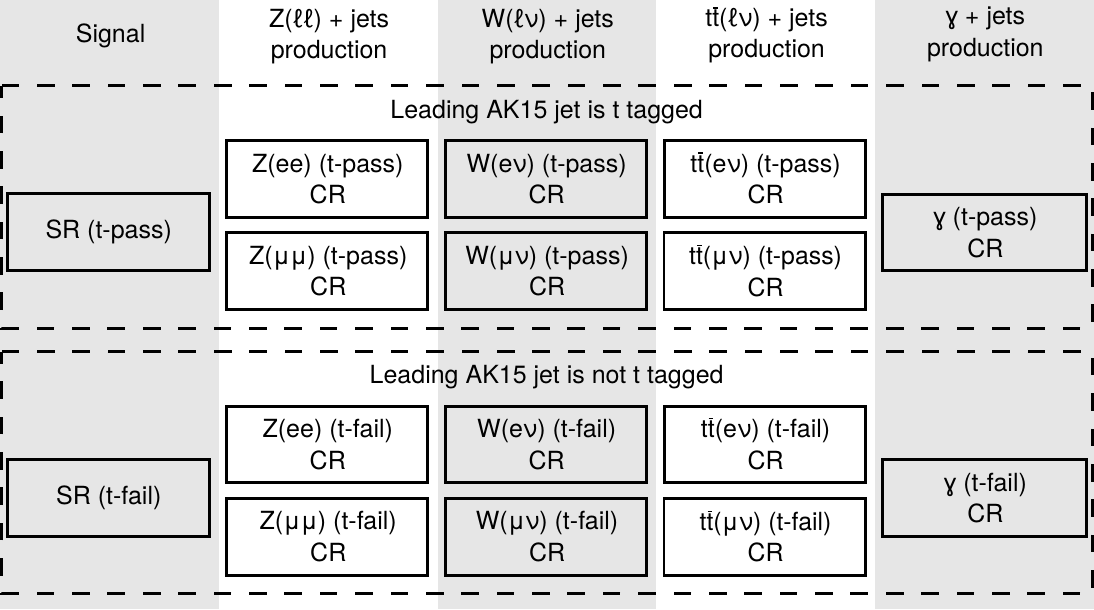}
  \caption{%
    Categorization of events into SRs and CRs, which are sensitive to specific processes, namely the mono-top signal, \vjets ($\PV = \PZ, \PW, \PGg$) processes, and \ttbar production.
    Each column contains categories that target the same process.
    For CRs with leptons in the final state, a version with electrons and a version with muons in the final states exist.
    Finally, in each category a split is performed based on whether the leading AK15 jet is \PQt tagged or not.
  }
  \label{fig:top_tagging:categories}
\end{figure}

A requirement on the minimal value of the discriminant in Eq.~\eqref{eq:top_quark_tagging:t_vs_qcd} is used to tag the leading AK15 jet as a \PQt jet or as a QCD jet.
The working point is chosen such that the misidentification rate of QCD jets being tagged as \PQt jets is 1\% in simulated \znunujets events.
In the SR and each CR, the \PQt tagging of the leading AK15 jet is used to split events into two samples, according to whether the leading AK15 jet in the event is tagged as a \PQt jet (\PQt-pass) or not (\PQt-fail).
An overview of all analysis categories after applying the \PQt tagging requirement is given in Fig.~\ref{fig:top_tagging:categories}.
It should be noted that a \PQt-pass and \PQt-fail category are created for all SRs and CRs introduced in Section~\ref{sec:event_selection}.
The SR (\PQt-fail) category is not sensitive to signal, but is used as a closure test for the modeling of SM processes.

The discriminator is operated at a working point where the \PQt quark identification efficiency in simulation varies from 45 to 80\% for AK15 jets that pass the \PQt tagging working point, depending on the AK15 jet \pt.
Top quark jets with higher \pt exhibit higher tagging efficiencies.
The inclusive \PQt tagging efficiency is approximately 67\%.
At this working point, the misidentification probability for AK15 jets of the QCD jet (\PQb jet) class is 1--2 (8--12)\% in simulation, depending on the AK15 jet \pt.

The \PQt tagging efficiency and misidentification rates are measured using three aggregated AK15 jet classes in the \ttbarlnu and \PGg CRs, respectively.
The jet classes are defined in simulation based on the number of quarks from the top quark decay that are matched within the AK15 jet cone, and whether there is \PQb flavor within the jet via ghost-hadron matching~\cite{Cacciari:2007fd}.
An AK15 jet that has at least two quarks from the hadronic top quark decay within its jet radius is defined as belonging to the \PQt jet class.
If an AK15 jet has at least one ghost hadron with \PQb flavor, which has been clustered into the reconstructed jet, but it does not fulfill the requirements of the \PQt jet class, it is assigned to the \PQb jet class.
AK15 jets that do not meet the requirements of either the \PQt jet nor the \PQb jet class form the QCD jet class.

The \ttbarlnu CRs represent an environment in which a large fraction of \PQt-tagged jets originates from a hadronic top quark decay.
The \PQt tagging efficiency is measured in a combined \ttbarlnu CR, which is merged from the \ttbarenu and \ttbarmunu CRs, defined in Section~\ref{sec:event_selection}.
The \PQt tagging efficiency in simulation is calculated by determining the fraction of AK15 jets in the \PQt jet class, which are also \PQt tagged.
In data, the number of all (\PQt-tagged) AK15 jets in the \PQt class is estimated by subtracting the number of all (\PQt-tagged) AK15 jets from the \PQb jet and QCD jet classes, both obtained from simulation.
The ratio of the \PQt-tagged AK15 jets to the total number of AK15 jets from the \PQt jet class is the efficiency of correctly identifying \PQt jets.
The misidentification rate of \PQt tagged AK15 jets that originate from QCD processes is determined in a similar manner in the \PGg CRs.
The \PQt tagging efficiency and the QCD misidentification rate are measured as functions of the leading AK15 jet \pt.
To correct for residual differences of the \PQt tagging efficiency and the misidentification rate in data and simulation, the ratio of the efficiencies is used as a scale factor.
It is applied to simulation as a per-event weight based on the properties of the leading AK15 jet.
The \PQt tagging scale factor between data and simulation ranges between 0.6 and 1, where 1 is equivalent to the behavior of the \PQt tagging in simulation being perfectly consistent with that on data.
Variations associated with systematic uncertainties from Section~\ref{sec:systematic_uncertainties} are propagated to the measurement of the scale factors.

In simulation, around 5\% of the leading AK15 jets in the \PGg CR, and around 10\% of the leading AK15 jets in the \ttbarlnu CRs, are assigned to the \PQb jet class.
Additionally, in the \PQt-pass SR, approximately 50\% of the \PQt-tagged AK15 jets originate from QCD production.
The second-largest fraction (25--35\%) arises from top quark-initiated jets.
Mistagged \PQb jets constitute the third-largest group, with 15--20\% of the \PQt-tagged jets originating from bottom quarks.
No misidentification rate is measured in data for these cases.
The ratio of the efficiencies in data and simulation is assumed to be unity, and a 50\% uncertainty is assigned to this ratio.
This uncertainty is also propagated to the \PQt tagging and QCD misidentification rate measurements as a systematic uncertainty.

\begin{figure}[!ht]
  \centering
  \includegraphics[width=0.49\textwidth]{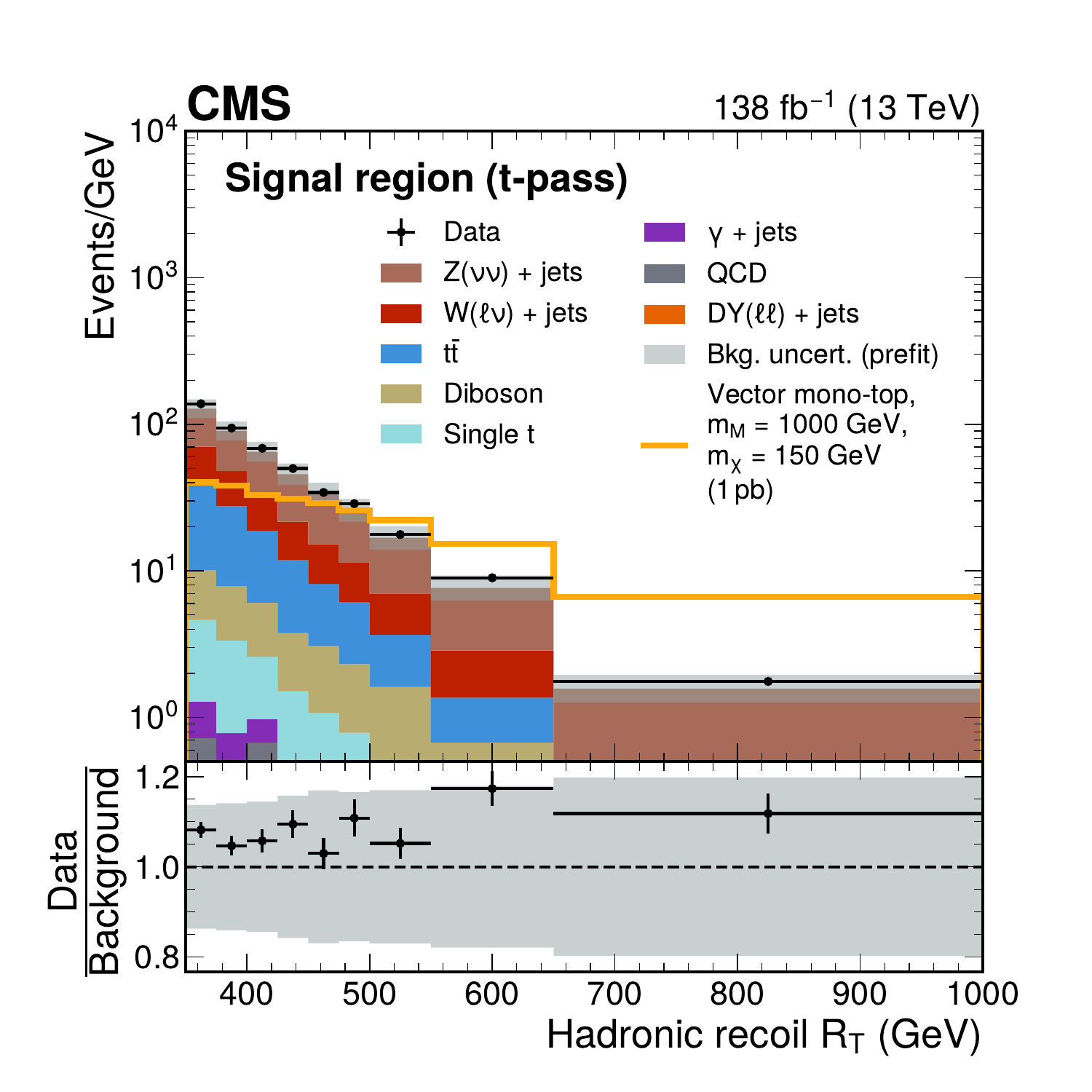}%
  \hfill%
  \includegraphics[width=0.49\textwidth]{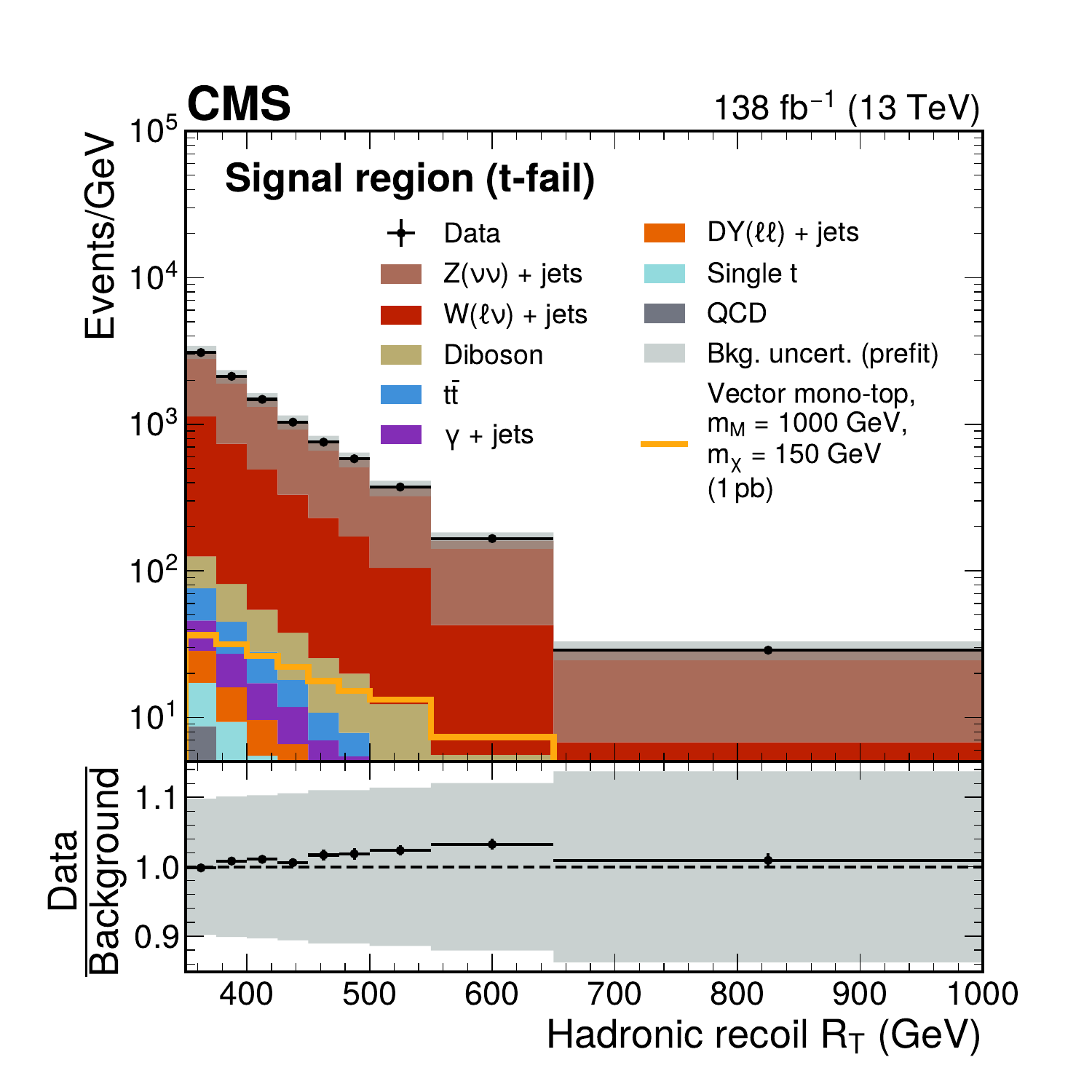}
  \caption{%
    Prefit distributions of the magnitude of the hadronic recoil \hadrecoil in the SR (\PQt-pass) and SR (\PQt-fail).
    The last bin of each distribution also contains events with $\hadrecoil > 1000\GeV$.
    The distributions of background processes from simulation are stacked together.
    All nominal corrections are applied to the simulated SM processes.
    A representative mono-top signal (vector coupling scenario) with a mediator mass of 1\TeV, a DM candidate mass of 150\GeV, and a cross section of 1\pb is overlaid as an orange line.
    The gray band represents the statistical and unconstrained systematic uncertainties in the simulated event yields.
  }
  \label{fig:prefit_recoil_pass_fail}
\end{figure}

For the SR (\PQt-pass) and the SR (\PQt-fail), the resulting distributions of the hadronic recoil after applying the split according to the \PQt tagging discriminant are shown in Fig.~\ref{fig:prefit_recoil_pass_fail}.
All background distributions are extracted from simulation.
The SR (\PQt-pass) region is enriched with signal events as well as background events containing hadronically decaying top quarks, \eg, \ttbar events.
However, \znunujets and \wlnujets events still are the largest background contribution because of their large cross sections, limited detector efficiency, and misidentification of purely QCD-initiated jets as top jets.
The SR (\PQt-fail) region almost completely consists of \znunujets and \wlnujets events.
It is mainly used to constrain systematic uncertainties related to the aforementioned processes together with the other CRs of the analysis.

\section{Background estimation and signal extraction}
\label{sec:background_estimation}

The signal is extracted with a binned maximum likelihood fit of the predicted backgrounds and the signal model to the \hadrecoil distribution in data.
All SRs and CRs, \ie, all analysis regions that are depicted in Fig.~\ref{fig:top_tagging:categories}, and all data-taking periods are combined in the fit.
Systematic uncertainties are parameterized by nuisance parameters.
The statistical analysis is carried out with \textsc{Combine}~\cite{CMS:2024onh} based on the \textsc{RooFit}~\cite{Verkerke:2003ir} and \textsc{RooStats}~\cite{Moneta:2010pm} software packages.

The \znunujets production constitutes the largest background contribution in the SR.
In order to constrain its yield and the shape of its \hadrecoil distribution as much as possible, data in CRs are used.
However, it is not possible to create a \znunujets CR for events with large \hadrecoil and a high-\pt, \PQt-tagged AK15 jet as there is no additional event activity to separate these events from the SR.
Instead, processes similar to \znunujets, for which dedicated CRs can be constructed, such as \wjets, \zjets, and \gammajets, are used for this purpose.
In these processes, events with a vector boson or photon, as well as a high-\pt AK15 jet are produced in the phase space considered in this analysis, similar to \znunujets.

The estimation of \znunujets production in the SR is directly implemented into the formalism of the maximum likelihood fit.
The same method has previously been used in \cite{cite:monotop_had_CMS_2016}.
In each bin of the \hadrecoil distribution, the \znunujets bin yield in the SR is estimated as freely floating parameter \rZNuNu in the fit.
In order to constrain \rZNuNu, other \vjets process contributions in analysis regions of this search are expressed as the product $t_{\vjets}\rZNuNu$ of a transfer factor (TF) and \rZNuNu.
The TFs 
\begin{equation}
  \label{eq:background_estimation:tf}
  t_{\vjets} = \frac{n_{\vjets}^{\text{sim}}}{n_{\znunujets}^{\text{sim}}}
\end{equation}
are obtained from simulation, where $n_{\vjets}^{\text{sim}}$ and $n_{\znunujets}^{\text{sim}}$ are the expected bin yields of a targeted \vjets process in a specific analysis region and of \znunujets in the SR, respectively.
In total, six \vjets processes in the SR and CRs are expressed as the product of a TF and the freely floating estimate \rZNuNu, namely:
\begin{itemize}
  \item \wlnujets in the SR;
  \item \zlljets in the \zee and \zmumu CRs;
  \item \wlnujets in the \wenu and \wmunu CRs;
  \item \wlnujets in the \ttbarenu and \ttbarmunu CRs;
  \item \gammajets in the \PGg CR.
\end{itemize}
This causes \rZNuNu to be strongly constrained in the fit using data from the \vjets CRs.

Similarly, the \hadrecoil distribution of \ttbar production in the SR is estimated as freely floating parameter \rTTBar for each bin of the distribution.
Here, the contribution of \ttbar production to CRs is used to constrain \rTTBar:
\begin{itemize}
  \item \ttbar in the \ttbarenu and \ttbarmunu CRs;
  \item \ttbar in the \wenu and \wmunu CRs.
\end{itemize}

\begin{figure}[htp]
  \centering%
  \includegraphics[width=0.98\textwidth]{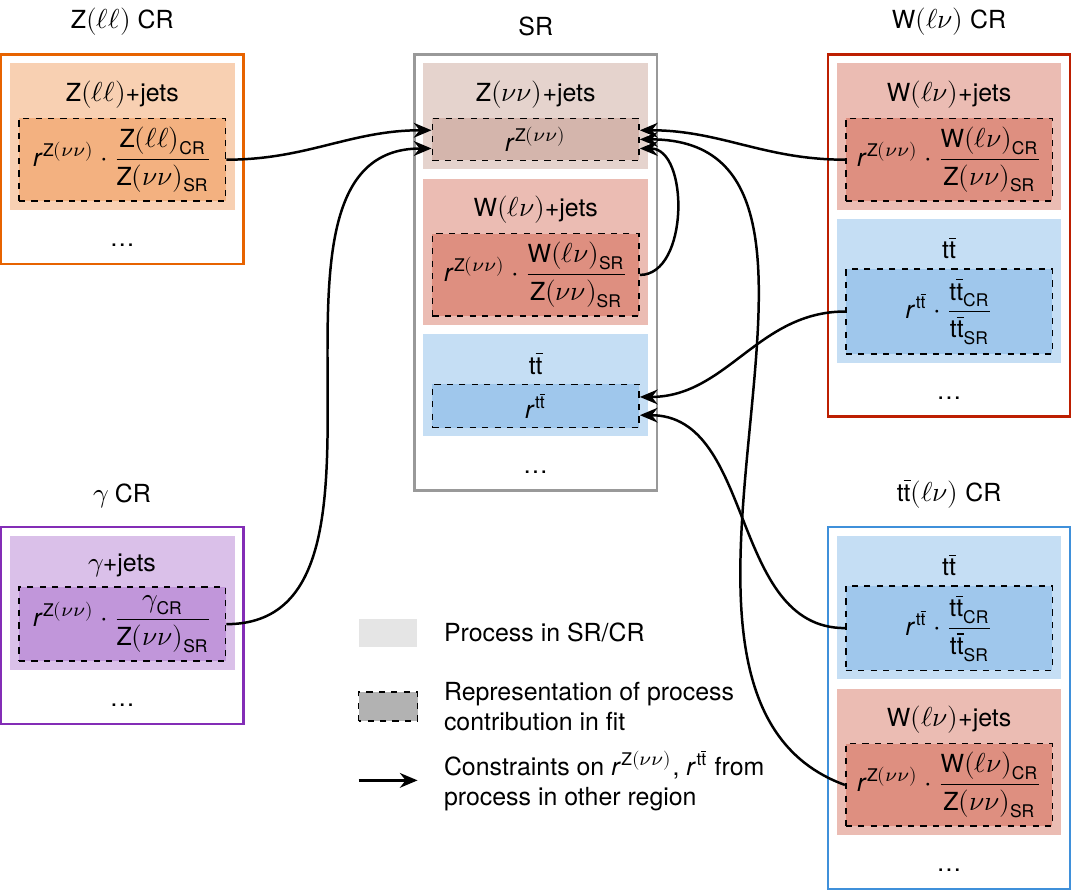}%
  \caption{%
    Statistical model used for the estimation of the major background processes for one bin of the \hadrecoil distribution.
    The contributions of \znunujets and \ttbar production in the SR are estimated with freely floating parameters \rZNuNu and \rTTBar.
    Constraints on \znunujets production are obtained by expressing similar processes in the SR and CRs as products of \rZNuNu and a TF, obtained from simulation.
    Concerning \ttbar production, the \ttbar processes in the \ttbarlnu and \wlnu CRs are expressed in terms of \rTTBar and a TF.
    All processes not depicted in this illustration are estimated using simulated events.
    Regions containing charged leptons are included twice in this model, once for electrons and once for muons.
    The model is implemented for the \PQt-pass and \PQt-fail regions separately.
  }%
  \label{fig:background_estimation:fit_model}%
\end{figure}

In both cases, the TF model is implemented independently for the \PQt-pass and the \PQt-fail regions.
Figure~\ref{fig:background_estimation:fit_model} shows the full TF model, as implemented in this search, for one bin of the \hadrecoil distribution.

The simulated yields in the TFs, as defined in Eq.~\eqref{eq:background_estimation:tf}, are subject to systematic uncertainties.
Their effects are propagated to the TFs by varying the corresponding simulated yields in the corresponding nuisance parameters.
All systematic uncertainties, listed in Section~\ref{sec:systematic_uncertainties}, are propagated to the TFs.
Since the TFs are ratios of similar processes in different CRs, several systematic uncertainties affect the numerator and denominator in the TF in a comparable way.
Because of this, these uncertainties are significantly reduced in the TFs.

To evaluate the importance of the CRs and TFs employed in the statistical model, fits in different scenarios with some CRs left out have been performed with the signal $\massPM = 2000\,\text{GeV}$, $\massPchiDM = 150\,\text{GeV}$ on Asimov pseudodata.
The results on the expected uncertainties in \rZNuNu in these fits are compared to a fit on a statistical model including all CRs.
In these fits, the \rTTBar parameters have been fixed to the nominal yield of \ttbar in the corresponding bin from simulation in order to isolate the effect on the \znunujets estimation.
When excluding the \zll CRs, effectively removing TFs with the \zlljets process from the fit model, the uncertainty on \rZNuNu increases by $50\%$ in low-\hadrecoil and by $10\%$ in high-\hadrecoil bins.
When excluding the \wlnu and the \ttbar CRs, effectively removing TFs with the \wlnujets process in CRs, the uncertainty on \rZNuNu increases by $90\%$ in low-\hadrecoil and by $30\%$ in high-\hadrecoil bins.
The relative uncertainty of \rZNuNu is approximately constant across the \hadrecoil range in these two scenarios.
When the $\gamma$ CRs are left out of the fit, which removes the $\PGg$ TFs from the fit, the uncertainty increases by around 20\% in low-\hadrecoil and by around $60\%$ in high-\hadrecoil bins.
The relative uncertainty in \rZNuNu increases for larger \hadrecoil.
Hence, concerning the TF method, the $\PGg$ CRs can put additional constraints on \rZNuNu in the high-\hadrecoil region compared to the other processes involved in the TF method.

Bias studies have confirmed that no bias on the fitted signal cross section is induced using this statistical model.

The modeling of QCD multijet production is validated in the SR by comparing the \hadrecoil distribution obtained from simulation with a shape extracted from a data-driven approach.
For each SR, a dedicated, mutually exclusive determination region (DR) enriched in QCD multi-jet events is built by applying all selection criteria for the SR except for the requirements on $\Delta\phi$(AK4 jets, \hadrecoilvec) and $\Delta\phi$(AK15 jet, \hadrecoilvec).
Events in the DRs must have at least one AK4 jet with $\Delta\phi\text{(AK4 jet, \hadrecoilvec)} < 0.5$, and the requirement of $\Delta\phi\text{(AK15 jet, \hadrecoilvec)} > 1.5$ is lifted.
The \hadrecoil distribution of QCD multijet production in the DR is estimated by subtracting the background yield of remaining processes, obtained from simulation, from the data yield in each bin of the \hadrecoil distribution.
The derived \hadrecoil distribution of QCD multijet production in the DR is mapped to the SR by multiplying the bin yield in the DR by dedicated TFs derived from a simulation of QCD multijet production.
The TFs are computed in each bin of the \hadrecoil distribution by dividing the predicted QCD multijet yield in the SR by the predicted QCD multijet yield in the corresponding DR.

The closure of the QCD multijet simulation is tested in the SRs by comparing the predicted \hadrecoil distributions, derived from simulation, and from an estimate based on control samples in data.
The difference between the predicted distributions from simulation and data ranges from 10 to 75\%, depending on \hadrecoil.
The simulation consistently underestimates the predicted number of QCD multijet events in the SR, compared to the distribution obtained from data.
The contamination of the SR with events from QCD multijet production is less than 1\%, compared to the total expected SM background, hence the contribution of QCD multijet production is expected to have a negligible impact on the total predicted SM background in the SRs.
In conclusion, simulated events are used to estimate the contribution of QCD multijet production.
To express the difference between the distributions obtained from simulation and from control samples in data, a normalization uncertainty of 100\% is assumed on the predicted yield of QCD multijet production in simulation.

\section{Systematic uncertainties}
\label{sec:systematic_uncertainties}

Several sources of experimental and theoretical uncertainty are considered in this analysis.
All systematic uncertainties are implemented as constrained nuisance parameters in the statistical model and are propagated to the final distributions and TFs after the simultaneous binned maximum likelihood fit used for signal extraction.
The systematic uncertainties can be grouped into two types: normalization and/or shape.
Normalization uncertainties only affect the total expected event yield.
Shape uncertainties can affect both the total number of events and the distribution of the final measured hadronic recoil.
Only systematic uncertainties that are directly related to the processes in Fig.~\ref{fig:background_estimation:fit_model} have an influence on the TFs.

In the following, the sources of systematic uncertainty considered in this analysis are described.
Unless otherwise stated, the systematic uncertainties are implemented as shape uncertainties.

\begin{itemize}
  \item \textit{Renormalization and factorization scales:}
    For the major backgrounds (\ttbar, \wjets, \zjets, \gammajets, and \singlet production) and the signal process the matrix element renormalization and factorization scales were varied up and down independently by a factor of 2 and 0.5, respectively.
    The uncertainties in the renormalization and factorization scale account for missing higher-order processes in the simulation.
    The up and down shifts are propagated to the simultaneous fit.
    The renormalization and factorization scale uncertainty for each process is uncorrelated during the fitting process.
    A 100\% correlation of these uncertainties across all data-taking periods is assumed.

  \item \textit{Parton distribution functions:}
    For the simulated samples, the PDFs are taken from NNPDF (v.~3.1)~\cite{NNPDF:2017mvq} and are used to evaluate the uncertainty associated with the choice of PDF.
    The 100 variations of the PDF set are determined using the eigenvector or Hessian~\cite{Buckley:2014ana,Butterworth:2015oua} approach and are then used to construct a 68\% confidence interval envelope around the nominal PDF value.
    This envelope is used as a shape uncertainty.
    The PDF uncertainties are assumed to be 100\% correlated for major background processes, which are mostly initiated by the same partons (namely a $\Pq\Pg$ or a $\Pg\Pg$ initial state).
    Processes with different initial states are treated as uncorrelated.
    The PDF uncertainty on the signal process is assumed to be uncorrelated with those on all background processes.
    In addition, the uncertainty in \alpS is propagated to the PDFs, which is treated as an additional shape uncertainty.
    This uncertainty is treated as 100\% correlated across all background processes.
    All of the presented uncertainties are 100\% correlated across all data-taking periods.

  \item \textit{Initial- and final-state radiation (ISR and FSR):}
    The parton shower scale in the simulation of major background samples was varied up and down by factors of 2 and 0.5, respectively, to evaluate the impact of \alpS on parton showers.
    The weights are obtained directly from simulation and used to generate input templates to the simultaneous fit.
    The ISR and FSR uncertainties are varied independently during the fitting procedure.
    These uncertainties are assumed to be 100\% correlated between the different data-taking periods for all the major backgrounds and uncorrelated between the signal prediction and background processes.
    ISR and FSR uncertainties are correlated across all data-taking periods.

  \item \textit{Top quark \pt reweighting:}
    The top quark \pt distribution of simulated \ttbar events is reweighted to NNLO(QCD)+NLO(EW) predictions~\cite{CMS:2016oae}.
    The simulation without reweighting and the doubled effect of the reweighting procedure are used as uncertainty of this method.
    This uncertainty is 100\% correlated across all data-taking periods.

  \item \textit{SM cross sections:}
    For the minor background processes of diboson and QCD multijet production, the uncertainties in the renormalization scale, the factorization scale, and the PDFs are replaced by an overall normalization uncertainty in the cross section predictions of these processes.
    The rate uncertainty in the total cross section of diboson production~\cite{Gehrmann:2014fva,Campbell:2011bn,Cascioli:2014yka} is 5\%.
    An uncertainty of 100\% is assigned to the total cross section of QCD multijet production, as described in Section~\ref{sec:background_estimation}.
    The cross section uncertainty is correlated across all taking periods for each process.

  \item \textit{\vjets electroweak correction:}
    The \wjets, \zjets, and \gammajets background models are reweighted to account for NLO EW corrections, as described in Ref.~\cite{Lindert:2017olm}.
    The theoretical uncertainties as functions of the vector boson \pt are provided by the authors and are applied to the \wjets, \zjets, and \gammajets samples.
    Altogether there are three sources of EW uncertainties: the unknown Sudakov logarithms beyond NNLO and/or next-to-leading logarithmic accuracy, the unknown hard (non-Sudakov) EW corrections beyond NLO, and the process-correlation effects.
    For the other two sources, the correlation across the processes is unknown, therefore no correlation is assumed for them.
    The uncertainties are correlated across all data-taking periods.

  \item \textit{Integrated luminosity:}
    The uncertainty in the measurement of the integrated luminosity is broken up into five separate parts~\cite{CMS:LUM-17-003, CMS:LUM-17-004, CMS:LUM-18-002}.
    There are three uncorrelated parts that are specific to the year in which the data were collected and two correlated parts.
    The first affects all three data-taking years, while the second only affects the last two data-taking years.
    They are implemented as normalization uncertainties of both the background and signal rates, with a range between 0.2 and 2.0\%.

  {\tolerance=800
  \item \textit{Lepton reconstruction and identification:}
    The reconstruction and identification efficiency for electrons, and the identification and isolation efficiency for muons in simulation is corrected to the performance observed in data from $\PZ/\PGg^\ast \to \Pell\Pell$ events, as a function of the lepton \pt and $\eta$.
    Each of the four corrections is derived independently.
    Due to the small size of the uncertainties, the statistical and systematic components of the uncertainty are combined and propagated to the simultaneous fit.
    The muon and electron uncertainties are uncorrelated from one another.
    Each of the four aforementioned uncertainties is correlated across CRs in the same data-taking period.
    As the efficiencies and associated uncertainties are dependent on detector configuration, they are uncorrelated across different data-taking periods.
  \par}

  \item \textit{Photon identification:}
    The identification efficiency for photons in simulation is corrected to the performance observed in data as a function of the photon's \pt and $\eta$.
    Due to the small size of the uncertainty the statistical and systematic components are combined and propagated to the final simultaneous fit.
    The uncertainty is uncorrelated across data-taking periods.

  \item \textit{Trigger efficiency:}
    The efficiency for electron, photon, and \ptmissnomu triggers is measured in data and residual differences to the efficiency in simulation are corrected~\cite{CMS:2024aqx}.
    The uncertainty associated with this correction is binned in \pt and $\eta$ for the lepton and photon trigger and in \ptmiss/\hadrecoil for the \ptmissnomu trigger and is propagated to the statistical model.
    The \ptmissnomu and single photon trigger efficiency is measured in an orthogonal sideband region and the single electron trigger efficiency is calculated using a tag-and-probe approach.
    The trigger efficiency uncertainty is correlated across CRs and SRs that utilize the same trigger.
    The uncertainty is uncorrelated across data-taking periods.

  \item \textit{\Lone prefiring:}
    During operation, a gradual timing shift of the ECAL was not properly propagated to the L1 trigger primitives (TP), resulting in a significant fraction of high-$\eta$ TP being mistakenly associated to the previous bunch crossing.
    Since L1 trigger rules forbid two consecutive bunch crossings to fire, a consequence of this is that events can self veto if a significant amount of ECAL energy is found in the region of $2 < \eta < 3$ (L1 ECAL prefiring).
    The effect is not described genuinely by the simulation but taken into account using event weights.
    This effect is most prominent in the 2016 and 2017 data-taking periods.
    The effects of the mistiming of the L1 ECAL trigger is propagated to the statistical model.
    The uncertainty is uncorrelated across the data-taking periods and applies only to the 2016 and 2017 data-taking years.

  \item \textit{Pileup:}
    The uncertainty in the distribution of the number of pileup interactions in the events is evaluated by varying the total inelastic \pp cross section in the simulation by $\pm$4.6\%~\cite{CMS:2018mlc}.
    To accommodate different beam and detector conditions, the uncertainty is uncorrelated across all data-taking periods

  \item \textit{Jet energy scale:}
    The uncertainty associated with the jet energy scale correction is estimated by shifting the jet energy scale up and down one standard deviation according to the source of uncertainty and then re-analyzing all the data with shifted jet energies~\cite{CMS:2016lmd, CMS-DP-2020-019, CMS-DP-2021-033}.
    Altogether eleven different sources of uncertainty in the jet energy scale are considered.
    The jet energy scale uncertainty is uncorrelated or correlated across data-taking periods depending on the source of uncertainty.
    A dedicated additional uncertainty is considered for the 2018 data set to account for any remaining effect of the HEM issue.

  \item \textit{Jet energy resolution:}
    The uncertainty in the jet energy resolution correction is evaluated by increasing or decreasing the difference between the reconstructed and parti\-cle-level jet energies~\cite{CMS:2016lmd, CMS-DP-2020-019, CMS-DP-2021-033}.
    The total uncertainty of this process is propagated to the final simultaneous fit and is uncorrelated across the data-taking periods.

  \item \textit{\ptmiss unclustered energy:}
    The uncertainty in the \ptmiss calculation from unclustered energy is estimated by varying each particle type by its estimated resolution.
    This uncertainty is uncorrelated across data-taking periods.

  \item \textit{Jet \PQb tagging:}
    The \PQb tagging scale factors, and their associated uncertainties, are dependent on the properties of the jet~\cite{BTV-16-002}.
    The uncertainty in the selection efficiency and misidentification rate are propagated to the statistical model.
    There is one selection and one misidentification uncertainty that are correlated across all data-taking periods and eight selection and misidentification uncertainties that
    are not correlated across the data-taking periods.

  \item \textit{Jet \PQt tagging:}
    The selection efficiencies of the \progParticleNet \PQt tagger~\cite{cite:particlenet} in simulation are corrected with scale factors to match the corresponding data efficiencies.
    The uncertainties depend on the type of AK15 jet (\PQt jet, QCD jet, or \PQb jet) and on its kinematic properties.
    The uncertainty in the scale factor of the \PQt tagging efficiency is 9--16\%.
    The uncertainty in the scale factor of the misidentification rate of an AK15 QCD jet is 12--18\%.
    As events with misidentified AK15 \PQb jets only represent a minor contribution, a scale factor of 1 and an uncertainty of 50\% on it is assumed.
    The misidentification rate uncertainties largely cancel in the TFs, which reduces the impact of these uncertainties in the fit presented in Section~\ref{sec:resultsandinterpretation}.
    The uncertainties in the \PQt tagging efficiency and misidentification rate are uncorrelated across data-taking periods.

  \item \textit{Limited size of simulation samples:}
    The limited size of the samples of simulated events results in statistical fluctuations on the nominal prediction.
    This uncertainty is taken into account with the simplified Barlow--Beeston method~\cite{Barlow:1993dm}.
    For bins with more than ten unweighted entries, a single Gaussian constrained nuisance parameter is introduced to each bin that scales the total yield.
    A Poisson probability distribution function is used for bins with fewer than ten entries.
\end{itemize}

\section{Results and interpretation}
\label{sec:resultsandinterpretation}

\begin{figure}[!ht]
  \centering
  \includegraphics[width=0.49\textwidth]{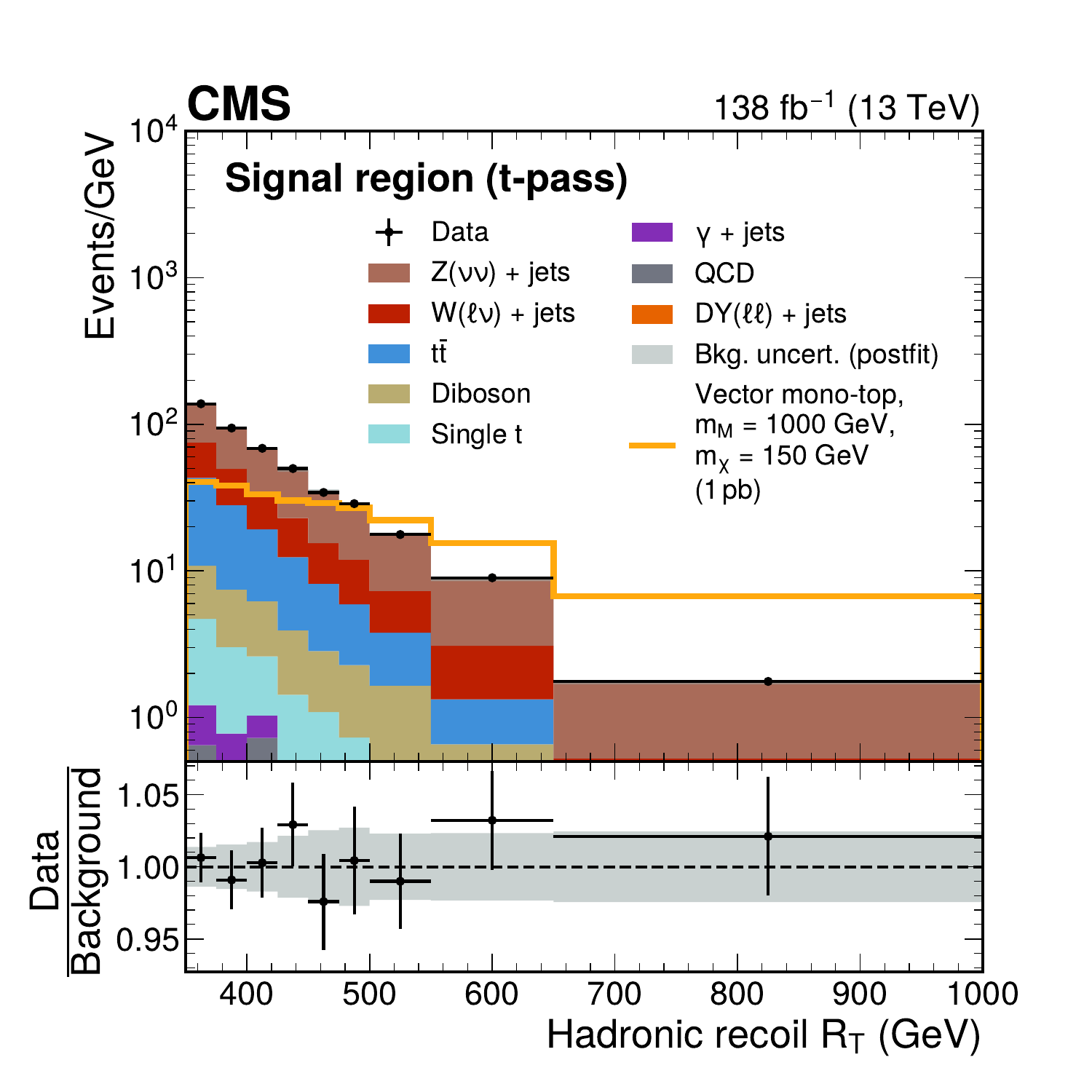}%
  \hfill%
  \includegraphics[width=0.49\textwidth]{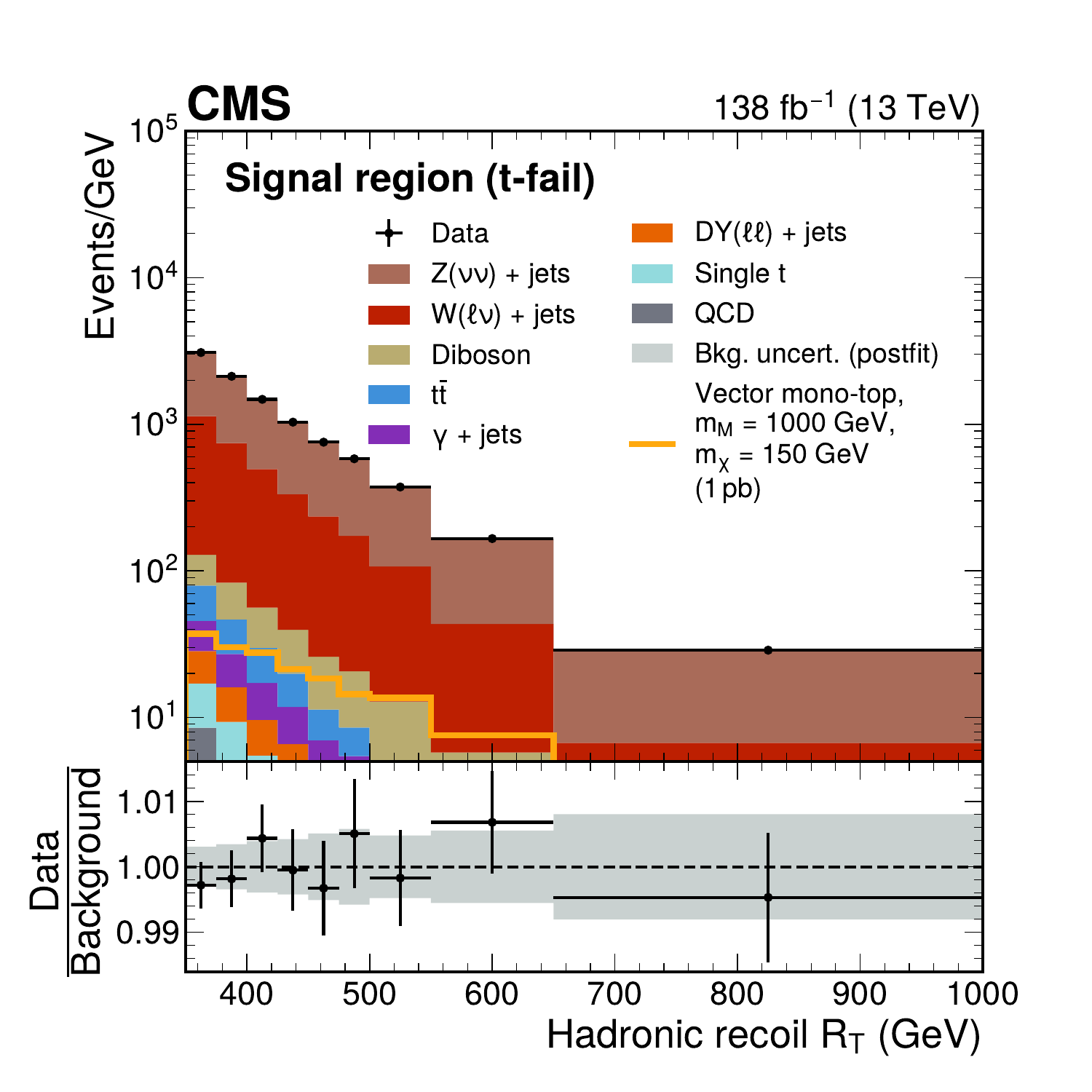}
  \caption{%
    Postfit distributions of the magnitude of the hadronic recoil \hadrecoil in the SR (\PQt-pass) and SR (\PQt-fail) after a fit of the background model to the data.
    The last bin of each distribution also contains events with $\hadrecoil > 1000\GeV$.
    The background processes are stacked together.
    A representative mono-top signal (vector coupling scenario) with a mediator mass of 1\TeV, a DM candidate mass of 150\GeV, and a cross section of 1\pb is overlaid as an orange line.
    The gray band represents the statistical and postfit systematic uncertainties in the predicted background yields after the fit.
  }%
  \label{fig:SR_postFitResults}
\end{figure}

The postfit distributions of the hadronic recoil in the SRs for a fit to the background-only hypothesis are shown in Fig.~\ref{fig:SR_postFitResults}.
The data are found to be in agreement with the SM prediction within uncertainties.
A goodness-of-fit test with the saturated model~\cite{Cousins:2018tiz} was performed to quantify the level of agreement of the data and the background-only postfit model that yielded a $p$-value of 0.72.
The postfit distributions in the CRs are shown in Appendix~\ref{sec:app_CRs_postFitResults}.

In addition, binned maximum likelihood fits of signal-plus-background hypotheses to data are performed.
Each tested signal-plus-background hypothesis is characterized by the choice of the unknown mediator mass and the DM candidate mass.
No significant deviation from the SM expectation is found in these fits.
Exclusion limits on the production strength of the mono-top signal, namely the product of the cross section and the branching fraction (\sigmaB) of $\pp \to \PQt{}\PM(\PchiDM\PAchiDM)$ production, are presented as functions of mediator masses between 200 and 2250\GeV and DM candidate masses between 1 and 1225\GeV, only considering on-shell decays of the mediator to the DM candidates.
The exclusion limits at 95\% confidence level (\CL) are calculated using the asymptotic approximation of the \CLs criterion~\cite{Cowan:2010js,cite:CLs_method_1,cite:CLs_method_2} using the LHC test statistic.
The observed and expected limits are calculated as function of \massPM for a constant $\massPchiDM = 150\GeV$.
For fixed values of \massPM, the limit on \sigmaB is assumed to be constant as function of \massPchiDM.
The signal shape of the hadronic recoil distribution does not depend on \massPchiDM if the DM production is on-shell, making such an extrapolation valid.

The impact of systematic uncertainties is checked for a signal with a vector mediator and with $\massPM = 2000\GeV$ and $\massPchiDM = 150\GeV$.
For this signal hypothesis, the expected exclusion limit roughly corresponds to the theory prediction of $\sigma\mathcal{B}$, promoting this hypothesis to a representative signal close to the expected exclusion contour.
An uncertainty breakdown of the fit shows that statistical and systematic uncertainties on the best fit $\sigma\mathcal{B}$ are at a similar level.
The expected exclusion limit is calculated with nuisance parameters related to a group of systematic uncertainties fixed.
The relative change of the limit compared to the fit on the full statistical model is used to quantify the impact of the systematic uncertainty on the results of the search.
The numerical results are summarized in Table~\ref{tab:results:uncertainty_impact}.
The largest impacts on the expected exclusion limit come from the uncertainties in the renormalization and factorization scales of the background processes, and from the limited size of the simulated samples.

\begin{table}[!ht]
  \centering
  \topcaption{
    Impact of systematic uncertainties on the expected exclusion limit for the vector mono-top signal with $\massPM = 2000\GeV$ and $\massPchiDM = 150\GeV$, quantified by the relative change in the expected exclusion limit when fixing the nuisance parameters related to a group of systematic uncertainties in the fit.
    The last row shows the impact on the exclusion limit if all nuisance parameters related to systematic uncertainties are fixed in the fit.
  }
  \label{tab:results:uncertainty_impact}
  \begin{tabular}{cc}
    \hline
    Uncertainty group                       & Change in exclusion limit on $\sigma\mathcal{B}$ \\
    \hline
    Renormalization and factorization scale & $9.7\%$                                          \\
    Limited size of simulated samples       & $6.9\%$                                          \\
    Remaining theory uncertainties          & $5.5\%$                                          \\
    Remaining experimental uncertainties    & $2.8\%$                                          \\
    \hline
    All systematic uncertainties            & $23.4\%$                                         \\
    \hline
  \end{tabular}
\end{table}

\begin{figure}[!p]
  \centering
  \includegraphics[width=0.95\textwidth]{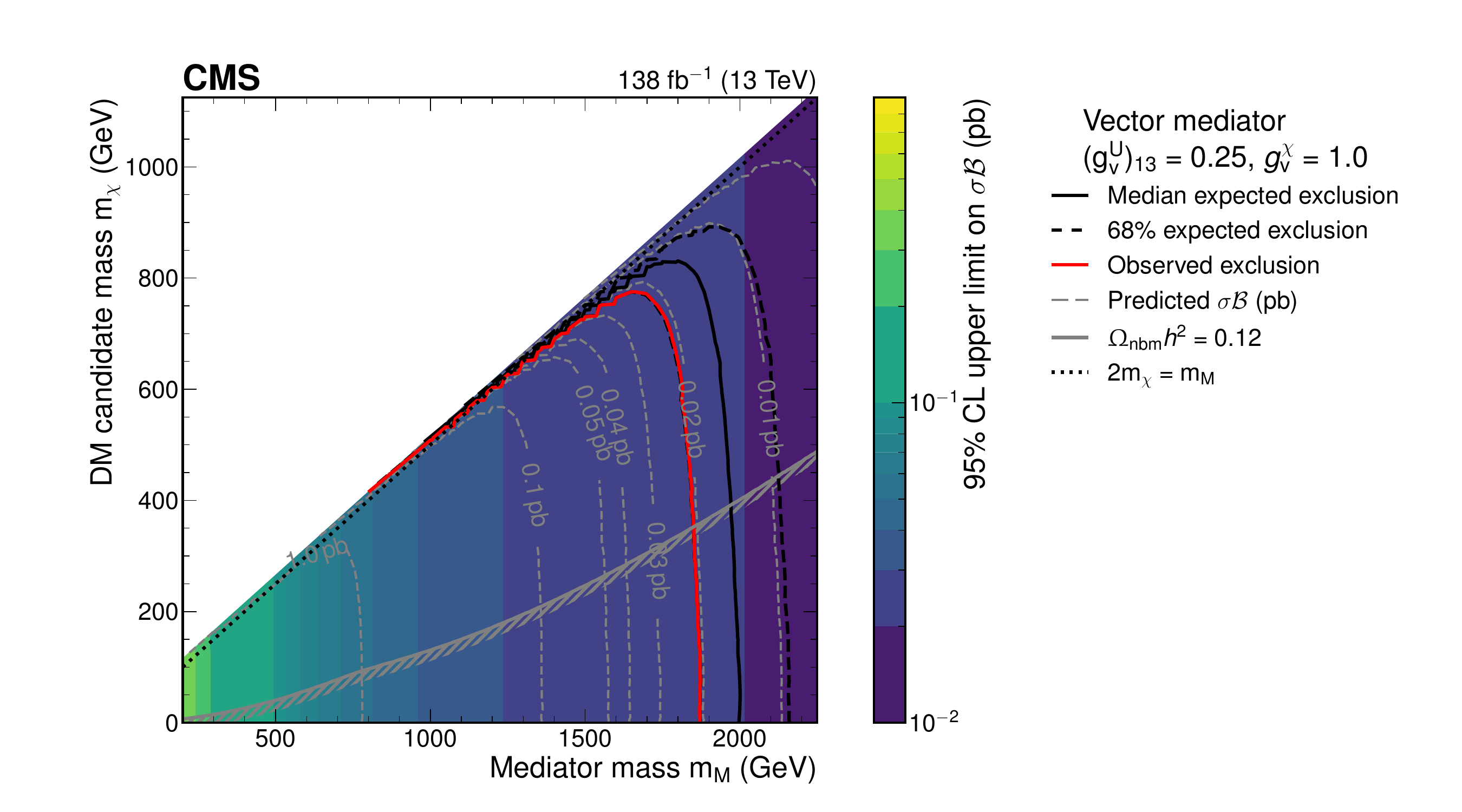}
  \\
  \includegraphics[width=0.95\textwidth]{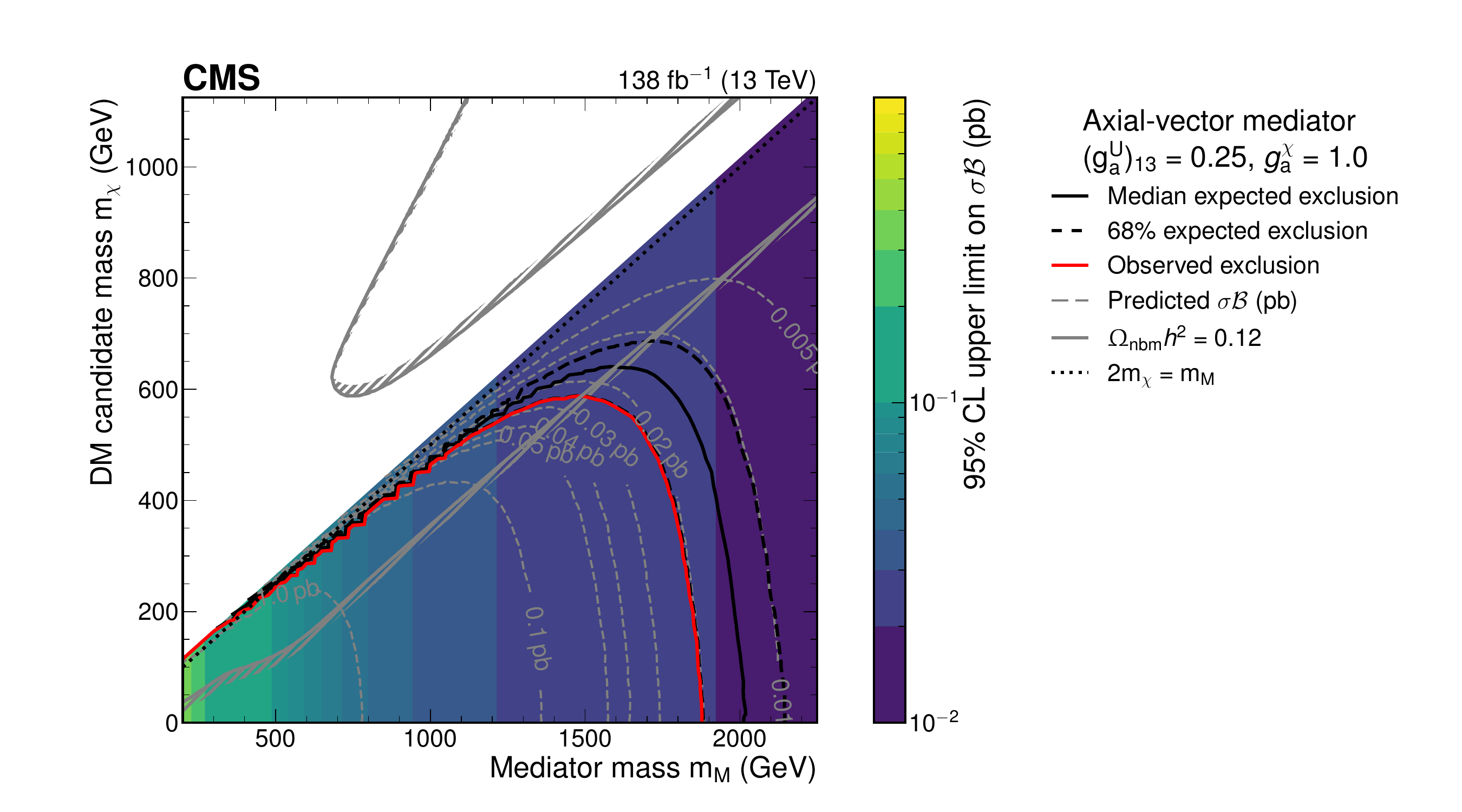}
  \caption{%
    Upper limits at 95\% \CL on \sigmaB of mono-top production presented in the two-dimensional plane spanned by the mediator and DM candidate masses for a mediator mass between 200 and 2250\GeV and a DM candidate mass between 1 and 1125\GeV only considering on-shell decays of the mediator to the DM candidates.
    The mediator has vector couplings to quarks and DM candidates in the upper plot and axial-vector couplings in the lower plot.
    The median expected exclusion range is indicated by a black solid line, demonstrating the search sensitivity of the analysis.
    The 68\% probability interval of the expected exclusion is shown in black dashed lines.
    Contours of theory predictions for constant values of \sigmaB are shown in gray dashed lines.
    The observed exclusion contour of mediator and DM candidate masses is represented by the red solid line.
    The exclusion contour obtained from measurements of the DM relic density \relicdensity by the Planck Collaboration is shown in the gray solid line.
  }%
  \label{fig:hadResults:all}
\end{figure}

Figure~\ref{fig:hadResults:all} shows the expected upper exclusion limits at 95\% \CL on the production strength \sigmaB of mono-top production as a function of the mediator and DM candidate mass assuming vector (upper part) and axial-vector (lower part) couplings exclusively.
For the vector (axial-vector) coupling scenario, $\gVUOneThree = 0.25$ and $\gVchi = 1$ ($\gAUOneThree = 0.25$ and $\gAchi = 1$) and all axial-vector (vector) couplings are set to zero.
For both, the vector and axial-vector coupling scenarios, an exclusion of mediator masses of up to 1.85\TeV is observed, where 2.0\TeV is expected.
For a mediator with pure vector (axial-vector) couplings, DM candidate masses of up to 750 (550)\GeV are excluded, where 850 (650)\GeV is expected.
The results of this search are comparable to a similar search~\cite{ATLAS:2024xne} performed by the ATLAS Collaboration using an alternative coupling scenario with couplings $\gVUOneThree = 0.5$ and $\gVchi = 1$, and a direct comparison can be made using the results in Appendix~\ref{sec:app_AltCouplingResults}.

The exclusion of mediator and DM candidate masses coming from DM relic constraints measured by the Planck Collaboration~\cite{cite:Planck_cosmo_params} is shown as a solid gray line.
For this line the measured value of the density of DM or nonbaryonic matter $\relicdensity = 0.12$ is used with $h$ representing the reduced Hubble constant.
Here, \relicdensity is the faction of the observable universe that is composed of nonrelativistic dark matter.
Mediator and DM candidate masses in the areas indicated by the hatched side of the solid gray contour in Fig.~\ref{fig:hadResults:all} result in higher relic densities than observed and are therefore excluded.
The relic densities for the corresponding mono-top models were calculated using \textsc{MadDM}~\cite{cite:MadDM_1,cite:MadDM_2}.

\section{Summary}
\label{sec:summary}

A search for dark matter (DM) produced in association with a single top quark via a flavor changing neutral current, referred to as nonresonant mono-top production, was presented.
The analysis was performed using data collected by the CMS experiment in 2016, 2017, and 2018 at the LHC at a center-of-mass-energy of \valueSqrtS, and corresponding to an integrated luminosity of \valueLumi.

The Lorentz boost of the top quark is exploited to cluster the products of the hadronic top quark decay into a large-radius jet.
Furthermore, a machine-learning-based discriminator is used to distinguish large-radius jets originating from hadronic top quark decays and large-radius jets produced purely through quantum chromodynamics processes.
A robust statistical model was built to determine the main backgrounds in the signal regions using data in dedicated control regions.
The distribution of the hadronic recoil in the signal and control regions is used to perform the statistical fit to the data.

The data are consistent with the background-only hypothesis, and no evidence for DM produced in association with a single top quark was found.
Limits at 95\% confidence level are calculated for the product of the signal production cross section and the branching fraction of the mediator decaying into DM candidates.
Limits were obtained for both a purely vector and a purely axial-vector mediator that couples to two DM candidates and to two standard model quarks: one from the first generation and another from the third.
The analysis excludes mediators with masses up to 1.85\TeV, where 2.0\TeV is expected, for both the vector and the axial-vector coupling scenarios.
Dark matter candidate masses below 750 (550)\GeV, where 850 (650)\GeV is expected, are excluded for the vector (axial-vector) coupling scenario.
In both cases, the exclusion limits are calculated for mediator masses greater than 200\GeV and DM candidate masses greater than 1\GeV.
The exclusion limit on the spin-1 mediator mass obtained in this search exceeds the previous CMS result on mono-top production~\cite{cite:monotop_had_CMS_2016} using the 2016 data set by 100\GeV.

\begin{acknowledgments}
  We congratulate our colleagues in the CERN accelerator departments for the excellent performance of the LHC and thank the technical and administrative staffs at CERN and at other CMS institutes for their contributions to the success of the CMS effort. In addition, we gratefully acknowledge the computing centers and personnel of the Worldwide LHC Computing Grid and other centers for delivering so effectively the computing infrastructure essential to our analyses. Finally, we acknowledge the enduring support for the construction and operation of the LHC, the CMS detector, and the supporting computing infrastructure provided by the following funding agencies: SC (Armenia), BMBWF and FWF (Austria); FNRS and FWO (Belgium); CNPq, CAPES, FAPERJ, FAPERGS, and FAPESP (Brazil); MES and BNSF (Bulgaria); CERN; CAS, MoST, and NSFC (China); MINCIENCIAS (Colombia); MSES and CSF (Croatia); RIF (Cyprus); SENESCYT (Ecuador); ERC PRG, RVTT3 and MoER TK202 (Estonia); Academy of Finland, MEC, and HIP (Finland); CEA and CNRS/IN2P3 (France); SRNSF (Georgia); BMBF, DFG, and HGF (Germany); GSRI (Greece); NKFIH (Hungary); DAE and DST (India); IPM (Iran); SFI (Ireland); INFN (Italy); MSIP and NRF (Republic of Korea); MES (Latvia); LMTLT (Lithuania); MOE and UM (Malaysia); BUAP, CINVESTAV, CONACYT, LNS, SEP, and UASLP-FAI (Mexico); MOS (Montenegro); MBIE (New Zealand); PAEC (Pakistan); MES and NSC (Poland); FCT (Portugal); MESTD (Serbia); MICIU/AEI and PCTI (Spain); MOSTR (Sri Lanka); Swiss Funding Agencies (Switzerland); MST (Taipei); MHESI and NSTDA (Thailand); TUBITAK and TENMAK (Turkey); NASU (Ukraine); STFC (United Kingdom); DOE and NSF (USA).

  \hyphenation{Rachada-pisek} Individuals have received support from the Marie-Curie program and the European Research Council and Horizon 2020 Grant, contract Nos.\ 675440, 724704, 752730, 758316, 765710, 824093, 101115353, 101002207, and COST Action CA16108 (European Union); the Leventis Foundation; the Alfred P.\ Sloan Foundation; the Alexander von Humboldt Foundation; the Science Committee, project no. 22rl-037 (Armenia); the Fonds pour la Formation \`a la Recherche dans l'Industrie et dans l'Agriculture (FRIA-Belgium); the Beijing Municipal Science \& Technology Commission, No. Z191100007219010 and Fundamental Research Funds for the Central Universities (China); the Ministry of Education, Youth and Sports (MEYS) of the Czech Republic; the Shota Rustaveli National Science Foundation, grant FR-22-985 (Georgia); the Deutsche Forschungsgemeinschaft (DFG), among others, under Germany's Excellence Strategy -- EXC 2121 ``Quantum Universe" -- 390833306, and under project number 400140256 - GRK2497; the Hellenic Foundation for Research and Innovation (HFRI), Project Number 2288 (Greece); the Hungarian Academy of Sciences, the New National Excellence Program - \'UNKP, the NKFIH research grants K 131991, K 133046, K 138136, K 143460, K 143477, K 146913, K 146914, K 147048, 2020-2.2.1-ED-2021-00181, TKP2021-NKTA-64, and 2021-4.1.2-NEMZ\_KI-2024-00036 (Hungary); the Council of Science and Industrial Research, India; ICSC -- National Research Center for High Performance Computing, Big Data and Quantum Computing and FAIR -- Future Artificial Intelligence Research, funded by the NextGenerationEU program (Italy); the Latvian Council of Science; the Ministry of Education and Science, project no. 2022/WK/14, and the National Science Center, contracts Opus 2021/41/B/ST2/01369 and 2021/43/B/ST2/01552 (Poland); the Funda\c{c}\~ao para a Ci\^encia e a Tecnologia, grant CEECIND/01334/2018 (Portugal); the National Priorities Research Program by Qatar National Research Fund; MICIU/AEI/10.13039/501100011033, ERDF/EU, "European Union NextGenerationEU/PRTR", and Programa Severo Ochoa del Principado de Asturias (Spain); the Chulalongkorn Academic into Its 2nd Century Project Advancement Project, and the National Science, Research and Innovation Fund via the Program Management Unit for Human Resources \& Institutional Development, Research and Innovation, grant B39G670016 (Thailand); the Kavli Foundation; the Nvidia Corporation; the SuperMicro Corporation; the Welch Foundation, contract C-1845; and the Weston Havens Foundation (USA).
\end{acknowledgments}

\bibliography{auto_generated}

\appendix

\section{Postfit distributions in control regions}
\label{sec:app_CRs_postFitResults}

\begin{figure}[!hp]
  \centering
  \includegraphics[width=0.49\textwidth]{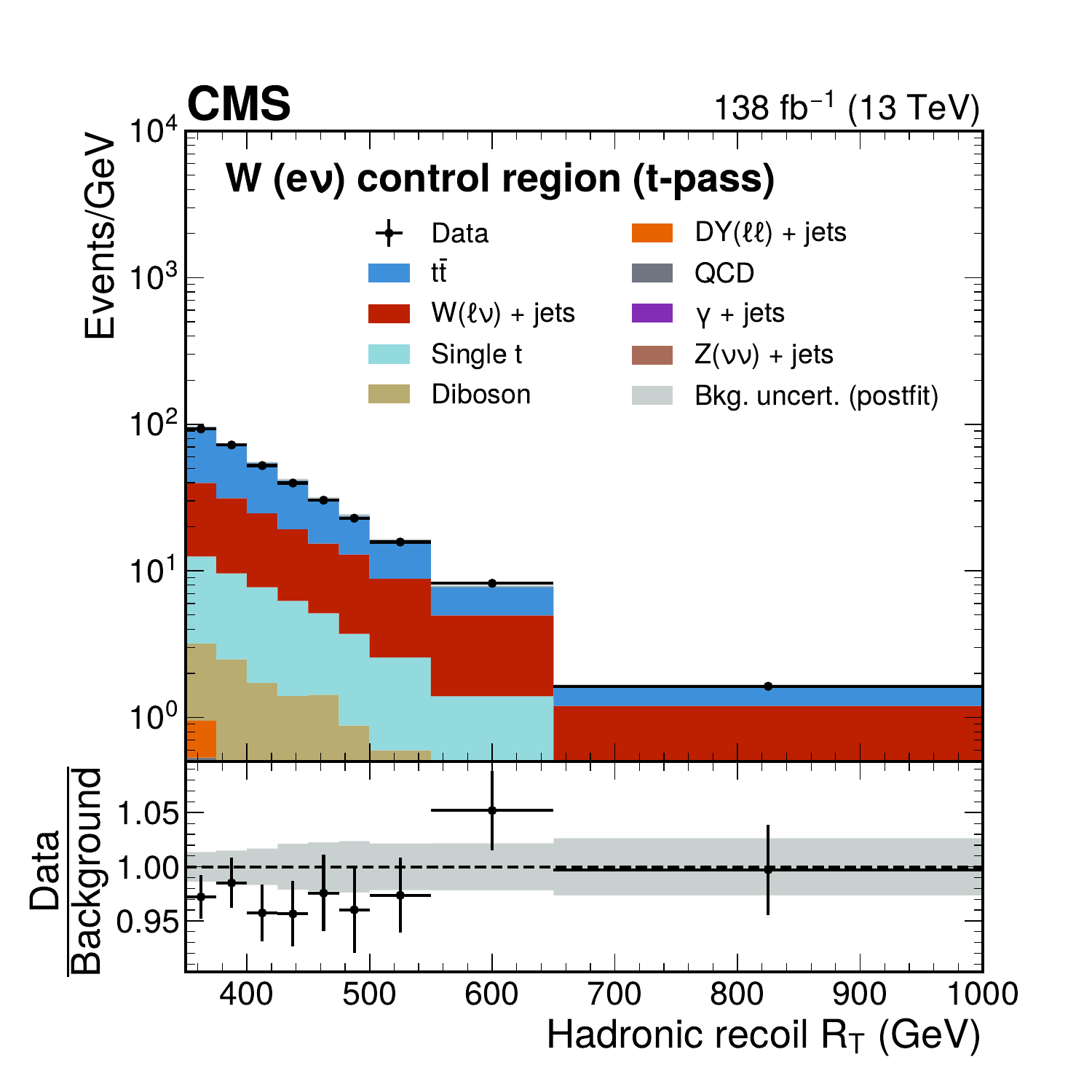}%
  \hfill%
  \includegraphics[width=0.49\textwidth]{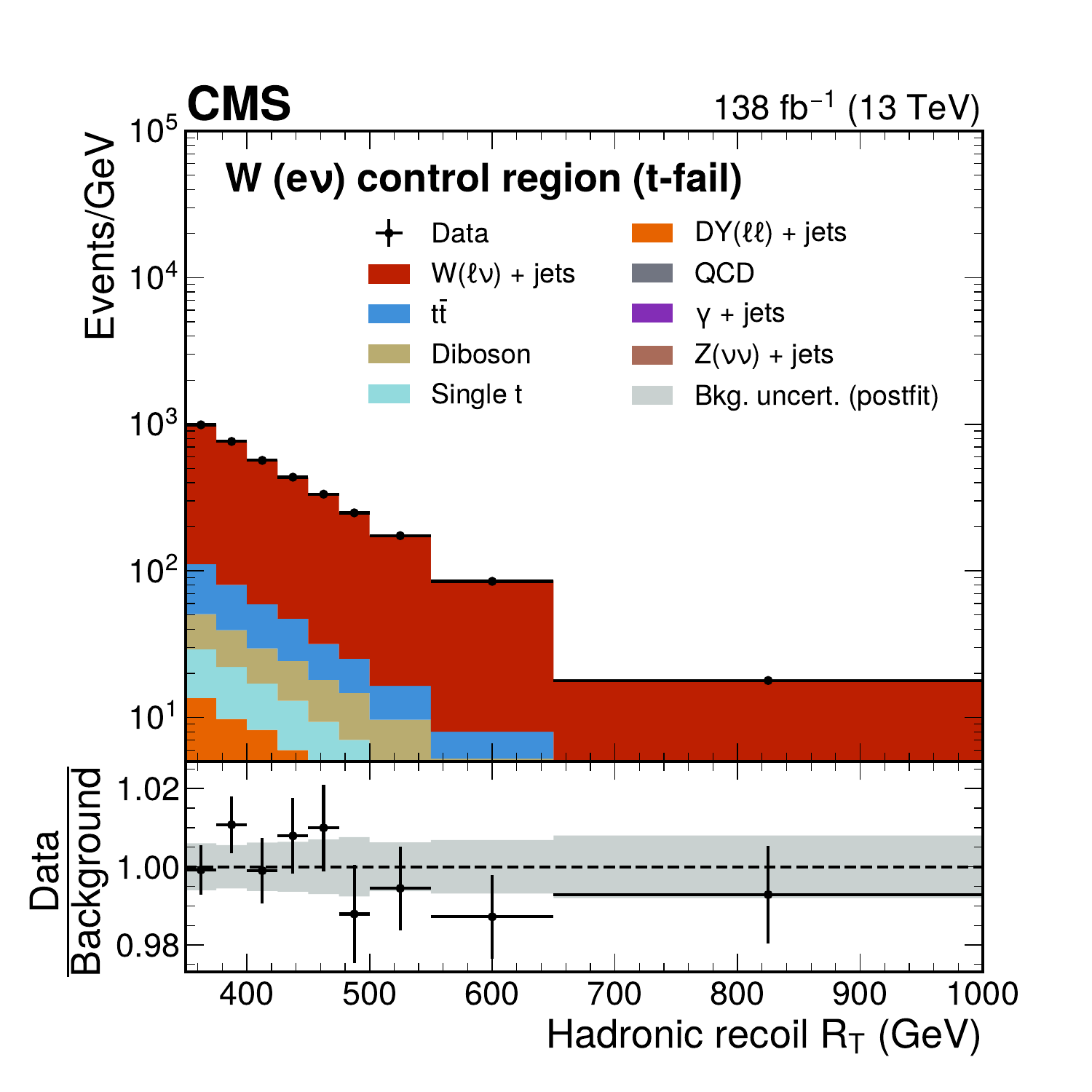}
  \caption{%
    Postfit distributions of the magnitude of the hadronic recoil \hadrecoil in the \wenu (\PQt-pass) and \wenu (\PQt-fail) CRs after a fit of the background model to the data.
    The last bin of each distribution also contains events with $\hadrecoil > 1000\GeV$.
    The background processes are stacked together.
    The gray band represents the statistical and postfit systematic uncertainties in the predicted background yields after the fit.
  }
  \label{fig:CR_W_electron_postFitResults}
\end{figure}

\begin{figure}[!hp]
  \centering
  \includegraphics[width=0.49\textwidth]{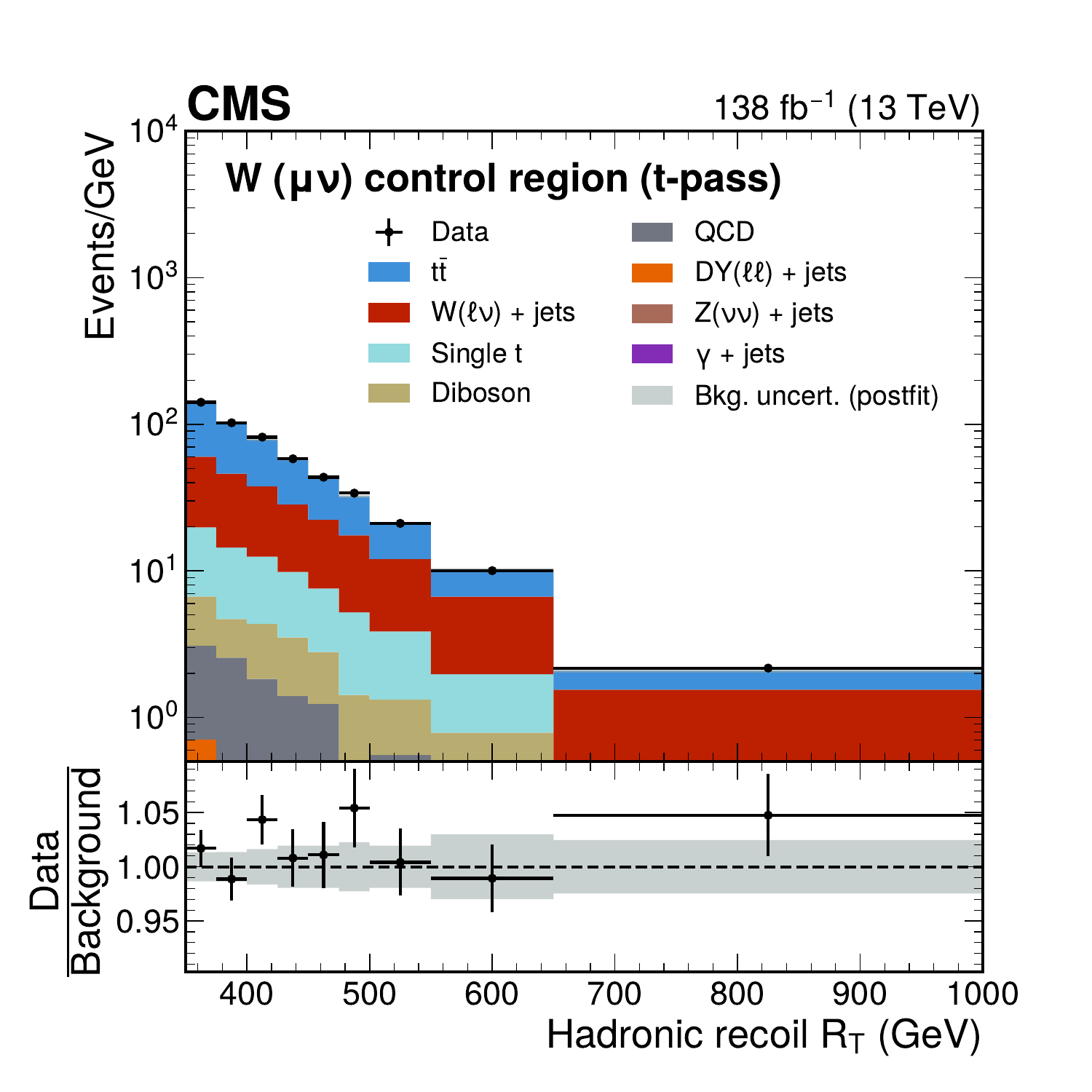}%
  \hfill%
  \includegraphics[width=0.49\textwidth]{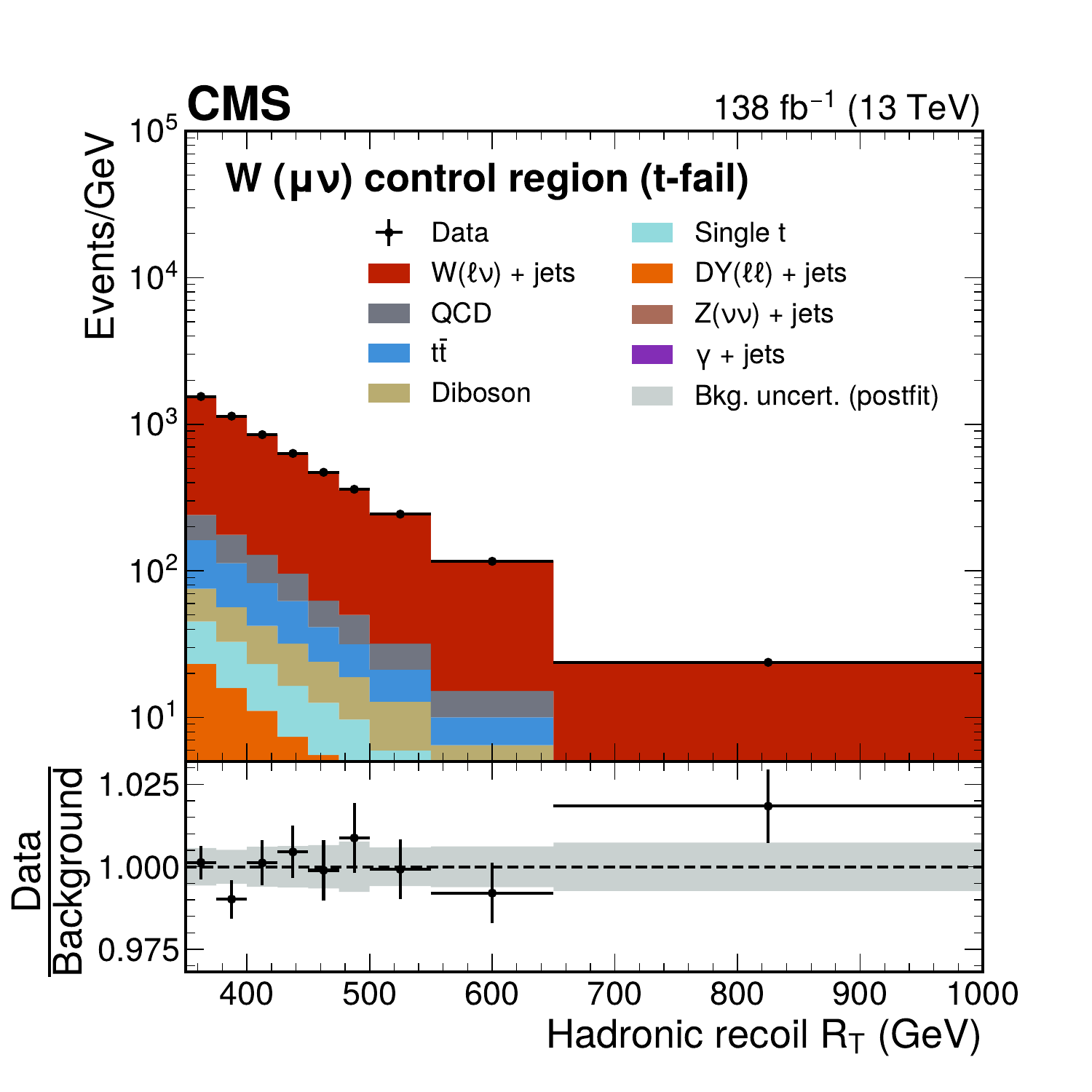}
  \caption{%
    Postfit distributions of the magnitude of the hadronic recoil \hadrecoil in the \wmunu (\PQt-pass) and \wmunu (\PQt-fail) CRs after a fit of the background model to the data.
    The last bin of each distribution also contains events with $\hadrecoil > 1000\GeV$.
    The background processes are stacked together.
    The gray band represents the statistical and postfit systematic uncertainties in the predicted background yields after the fit.
  }
  \label{fig:CR_W_muon_postFitResults}
\end{figure}

\begin{figure}[!hp]
  \centering
  \includegraphics[width=0.49\textwidth]{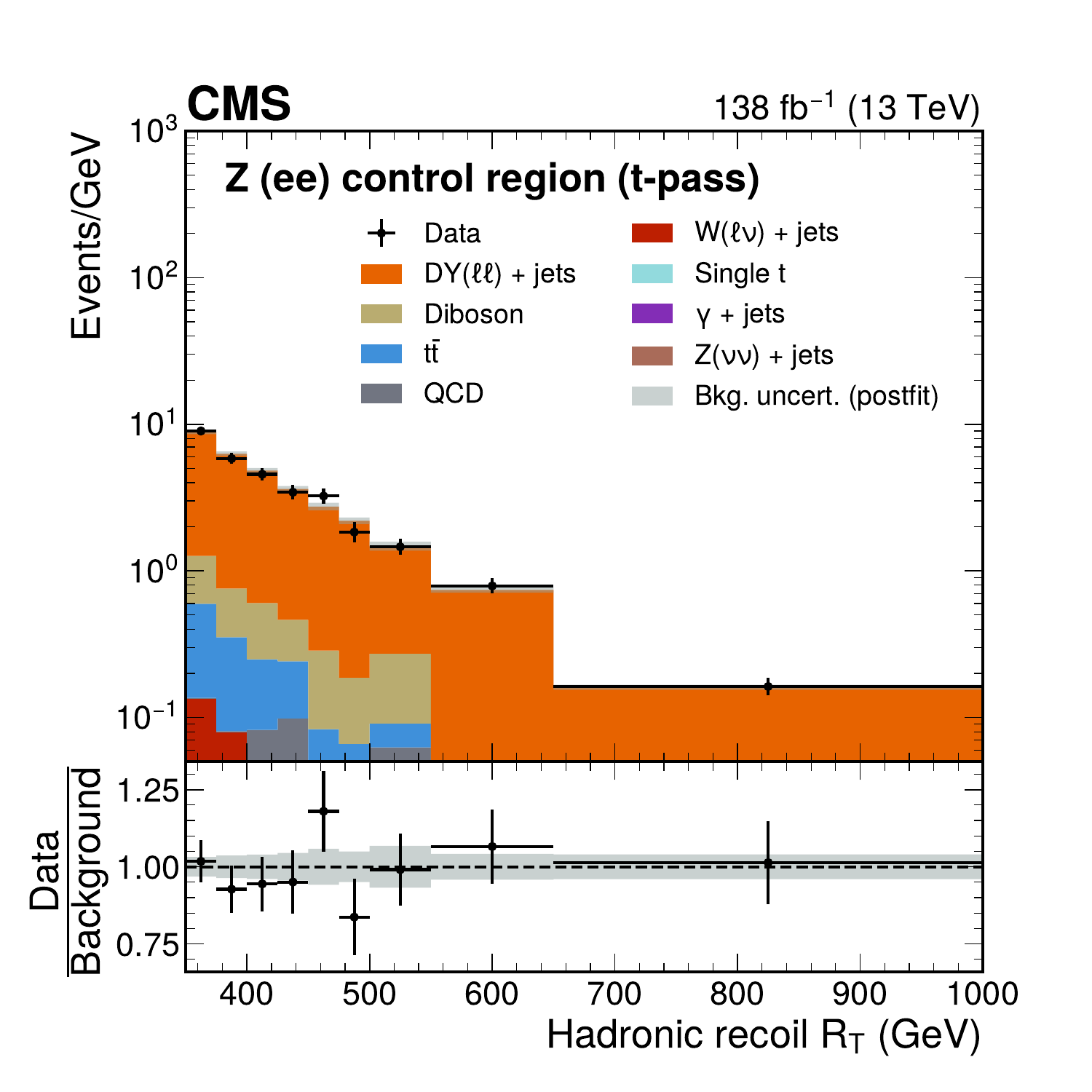}%
  \hfill%
  \includegraphics[width=0.49\textwidth]{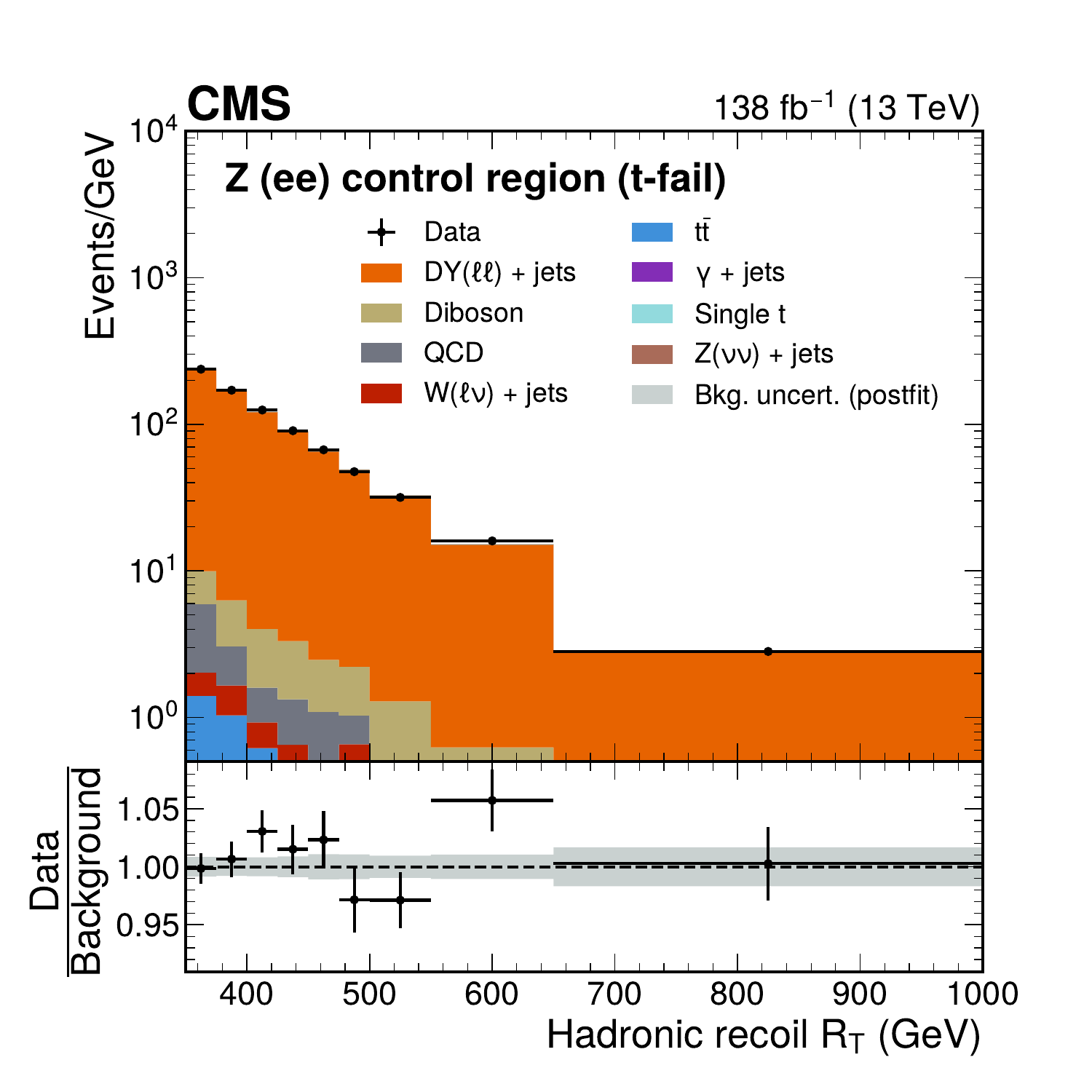}
  \caption{%
    Postfit distributions of the magnitude of the hadronic recoil \hadrecoil in the \zee (\PQt-pass) and \zee (\PQt-fail) CRs after a fit of the background model to the data.
    The last bin of each distribution also contains events with $\hadrecoil > 1000\GeV$.
    The background processes are stacked together.
    The gray band represents the statistical and postfit systematic uncertainties in the predicted background yields after the fit.
  }
  \label{fig:CR_Z_ElEl_postFitResults}
\end{figure}

\begin{figure}[!hp]
  \centering
  \includegraphics[width=0.49\textwidth]{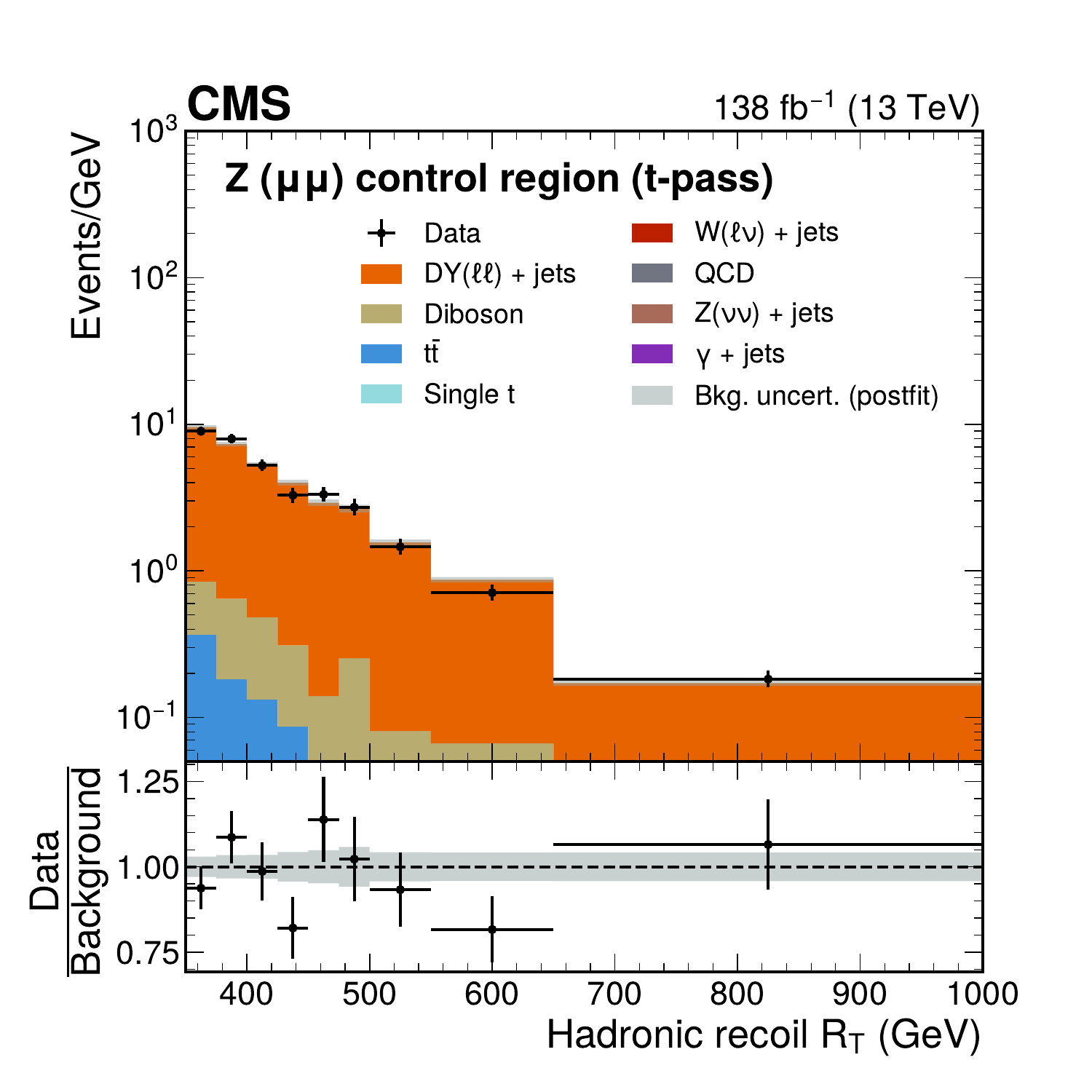}%
  \hfill%
  \includegraphics[width=0.49\textwidth]{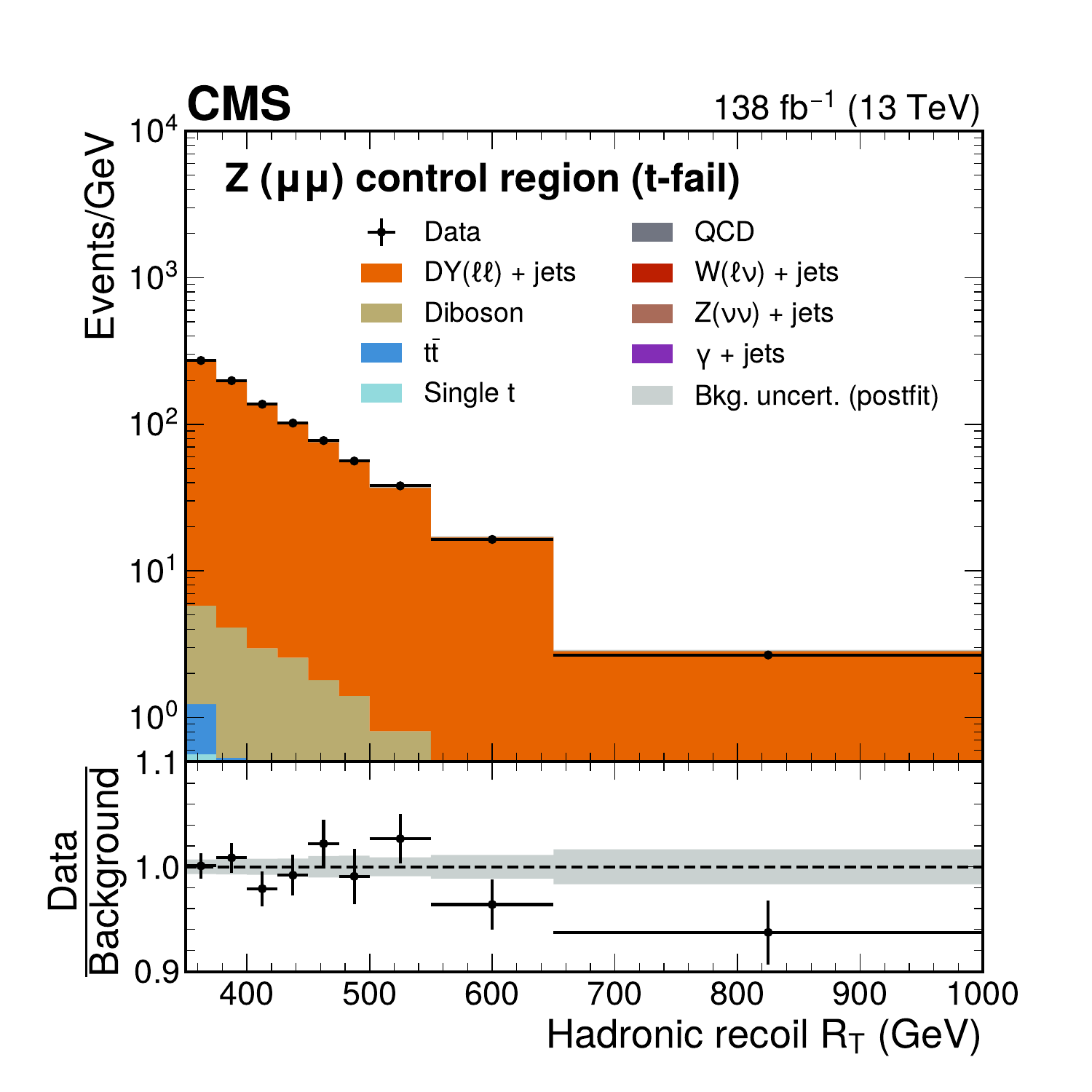}
  \caption{%
    Postfit distributions of the magnitude of the hadronic recoil \hadrecoil in the \zmumu (\PQt-pass) and \zmumu (\PQt-fail) CRs after a fit of the background model to the data.
    The last bin of each distribution also contains events with $\hadrecoil > 1000\GeV$.
    The background processes are stacked together.
    The gray band represents the statistical and postfit systematic uncertainties in the predicted background yields after the fit.
  }%
  \label{fig:CR_Z_MuMu_postFitResults}
\end{figure}

\begin{figure}[!hp]
  \centering
  \includegraphics[width=0.49\textwidth]{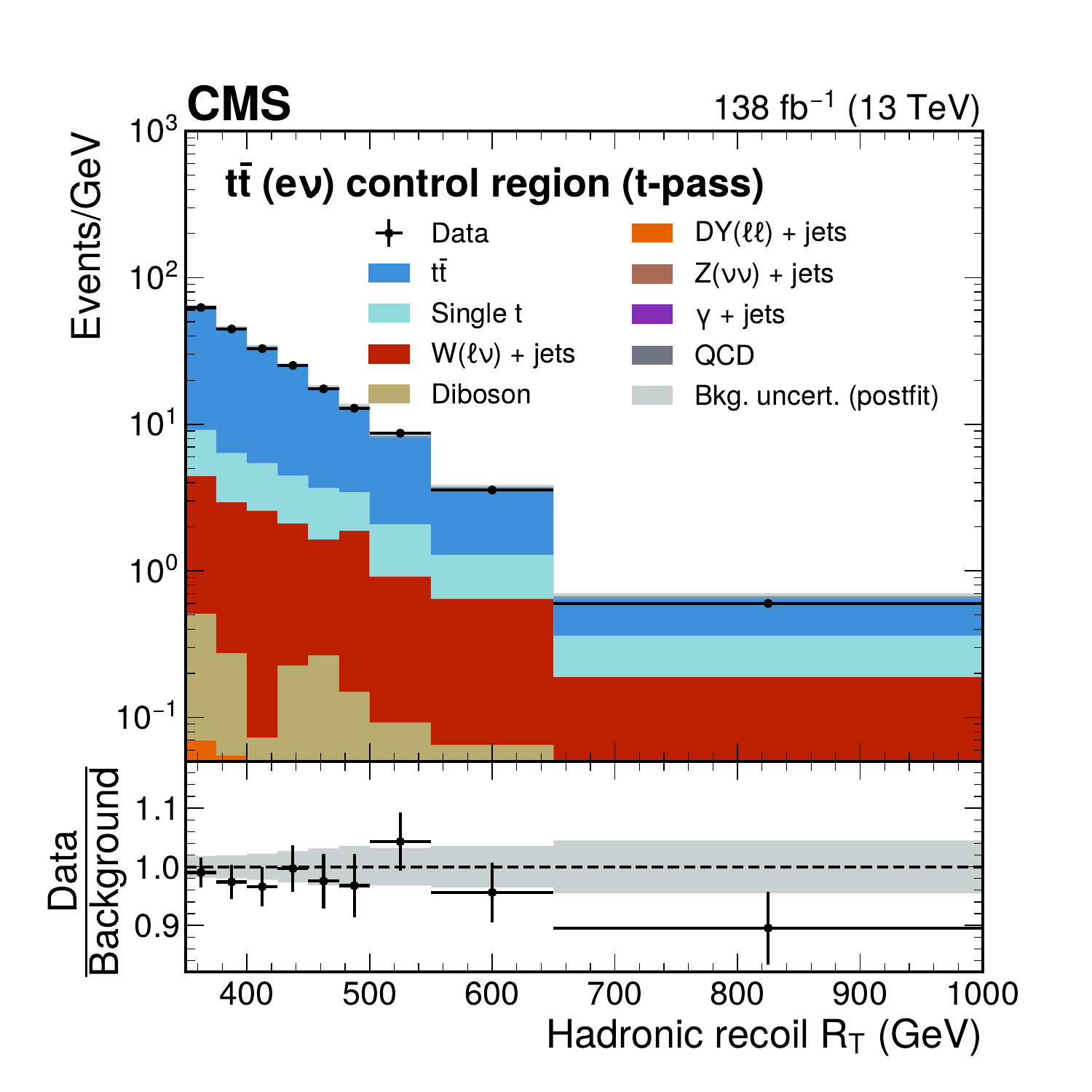}%
  \hfill%
  \includegraphics[width=0.49\textwidth]{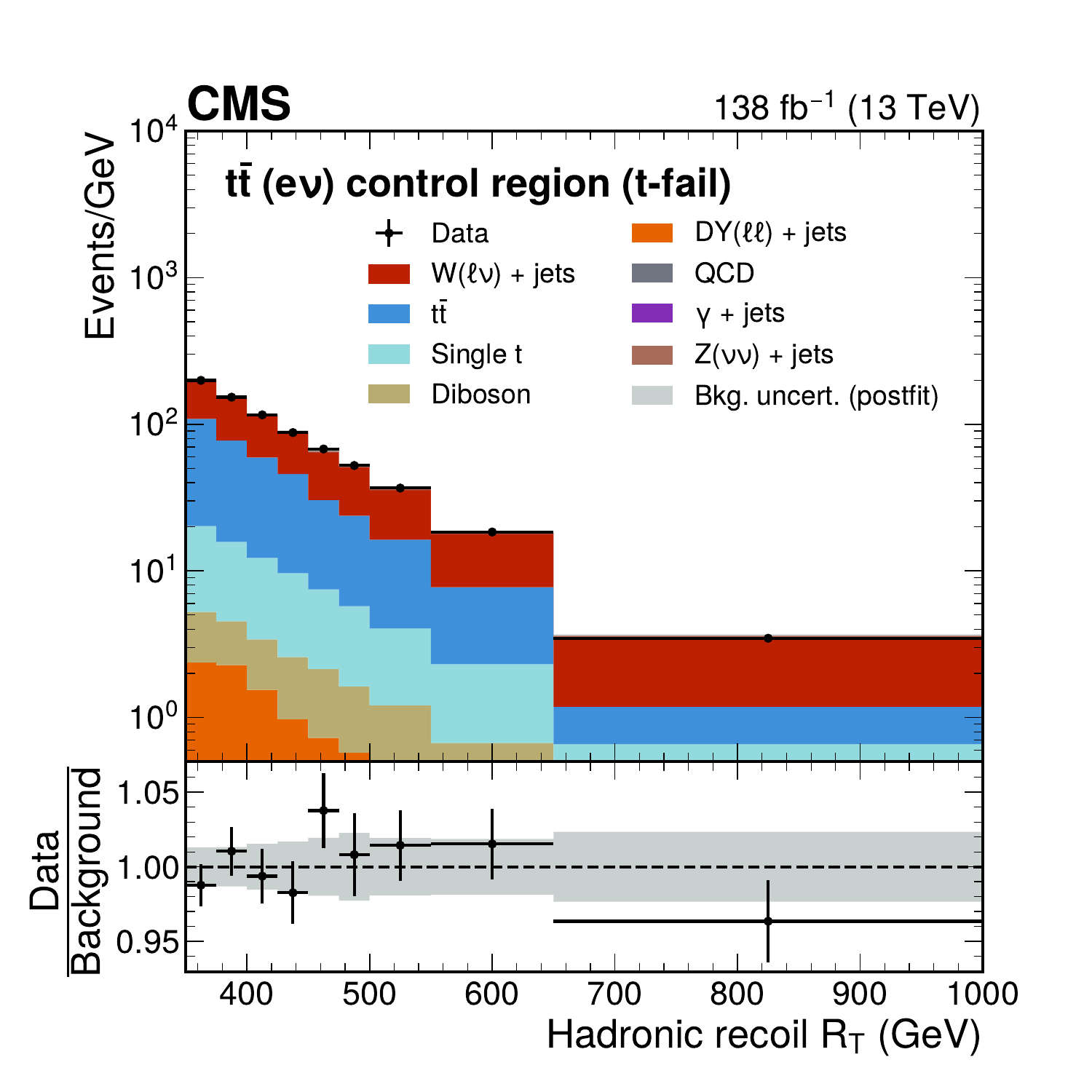}
  \caption{%
    Postfit distributions of the magnitude of the hadronic recoil \hadrecoil in the \ttbarenu (\PQt-pass) and \ttbarenu (\PQt-fail) CRs after a fit of the background model to the data.
    The last bin of each distribution also contains events with $\hadrecoil > 1000\GeV$.
    The background processes are stacked together.
    The gray band represents the statistical and postfit systematic uncertainties in the predicted background yields after the fit.
  }%
  \label{fig:CR_TT_electron_postFitResults}
\end{figure}

\begin{figure}[!hp]
  \centering
  \includegraphics[width=0.49\textwidth]{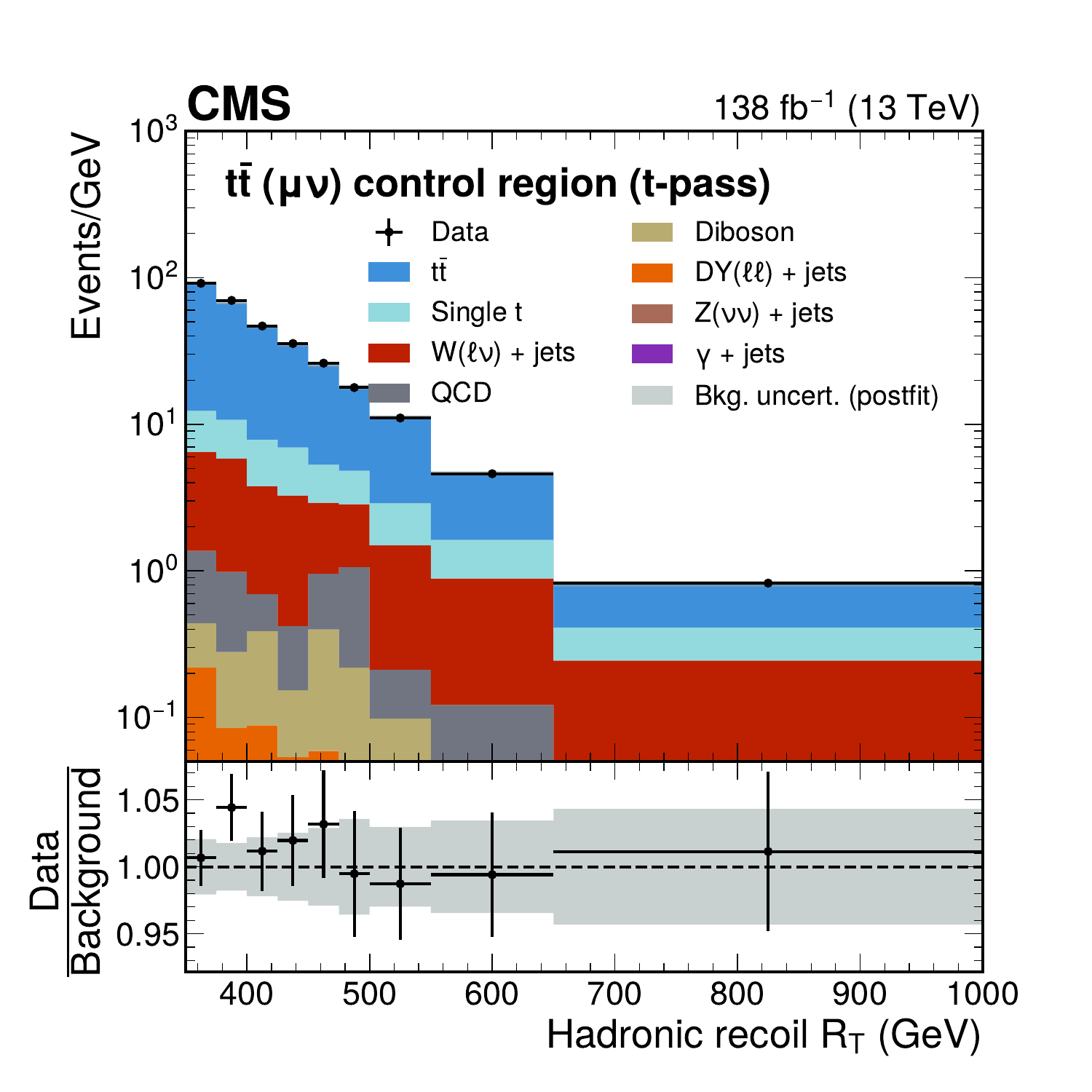}%
  \hfill%
  \includegraphics[width=0.49\textwidth]{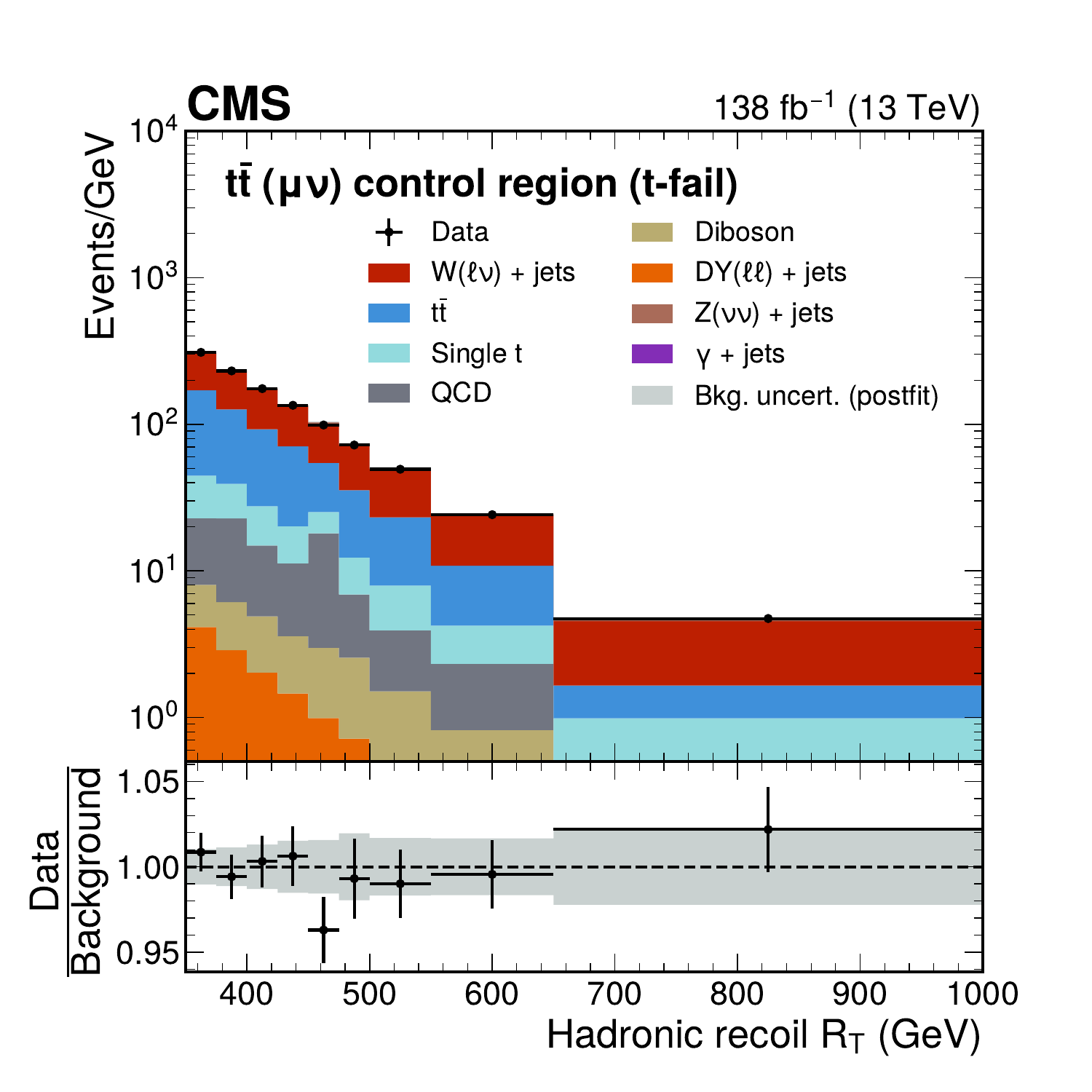}
  \caption{%
    Postfit distributions of the magnitude of the hadronic recoil \hadrecoil in the \ttbarmunu (\PQt-pass) and \ttbarmunu (\PQt-fail) CRs after a fit of the background model to the data.
    The last bin of each distribution also contains events with $\hadrecoil > 1000\GeV$.
    The background processes are stacked together.
    The gray band represents the statistical and postfit systematic uncertainties in the predicted background yields after the fit.
  }%
  \label{fig:CR_TT_muon_postFitResults}
\end{figure}

\begin{figure}[!hp]
  \centering
  \includegraphics[width=0.49\textwidth]{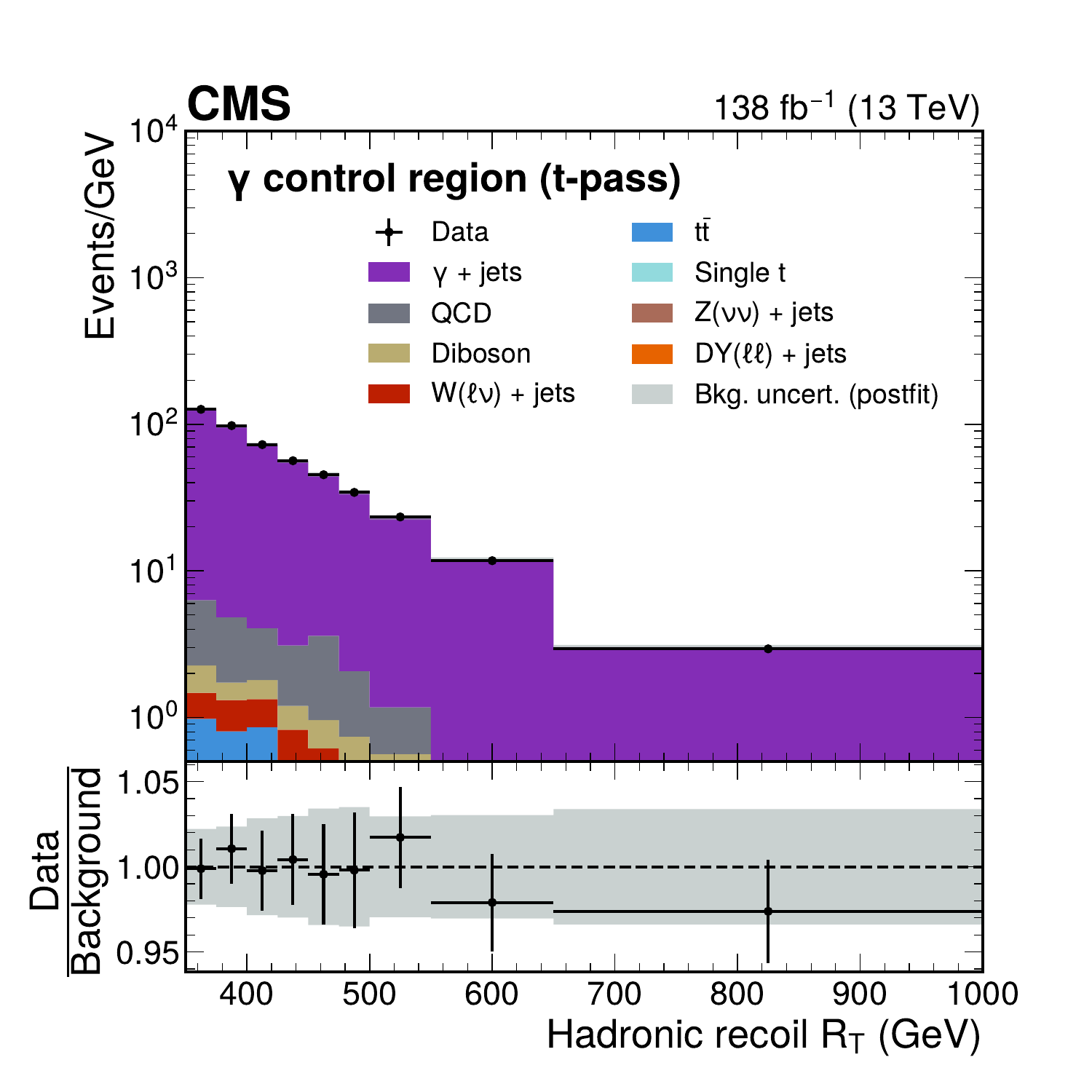}%
  \hfill%
  \includegraphics[width=0.49\textwidth]{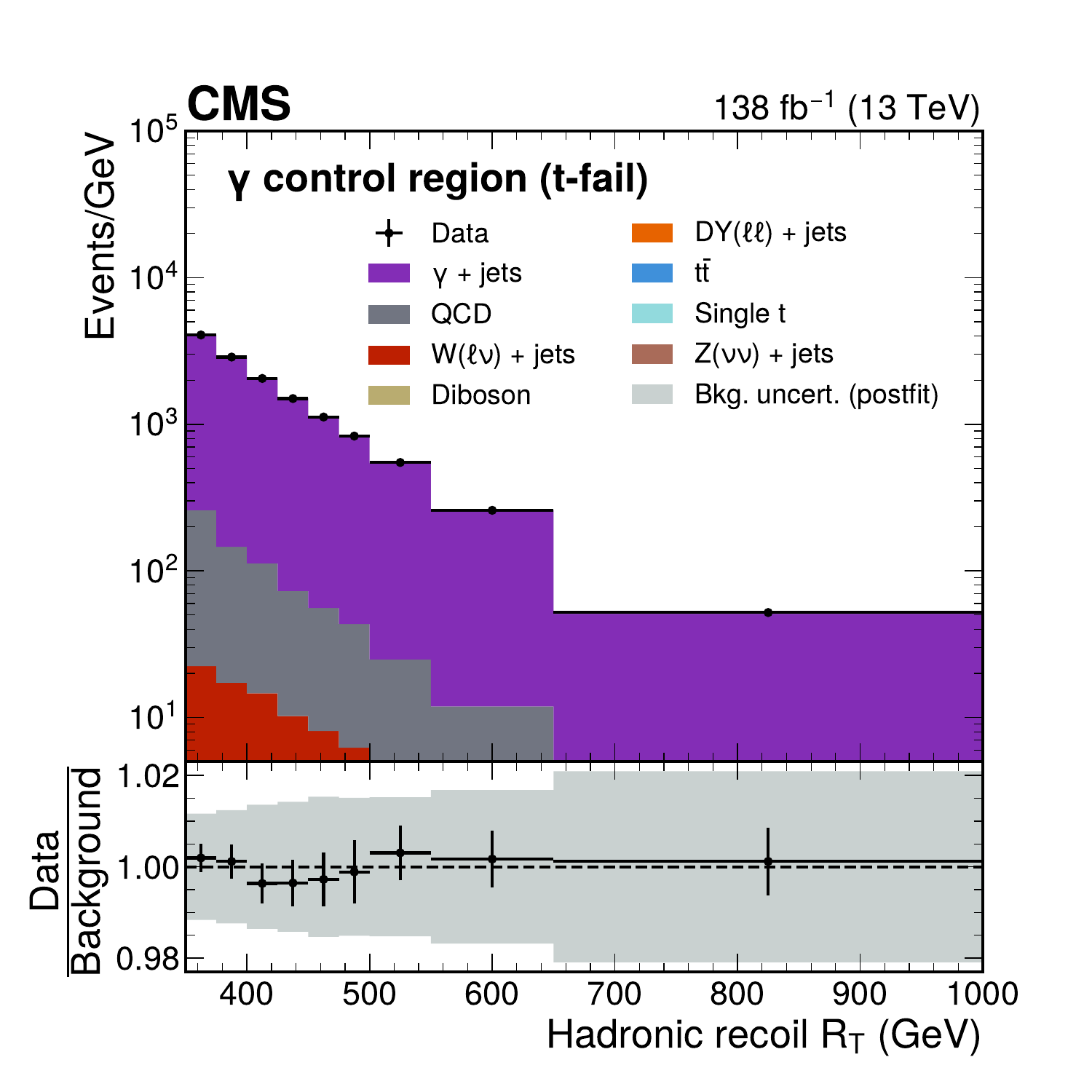}
  \caption{%
    Postfit distributions of the magnitude of the hadronic recoil \hadrecoil in the \PGg (\PQt-pass) and \PGg (\PQt-fail) CRs after a fit of the background model to the data.
    The last bin of each distribution also contains events with $\hadrecoil > 1000\GeV$.
    The background processes are stacked together.
    The gray band represents the statistical and postfit systematic uncertainties in the predicted background yields after the fit.
  }%
  \label{fig:CR_Gamma_postFitResults}
\end{figure}

\clearpage

\section{Exclusion limits for alternative choice of couplings}
\label{sec:app_AltCouplingResults}

\begin{figure}[!hp]
  \centering
  \includegraphics[width=0.95\textwidth]{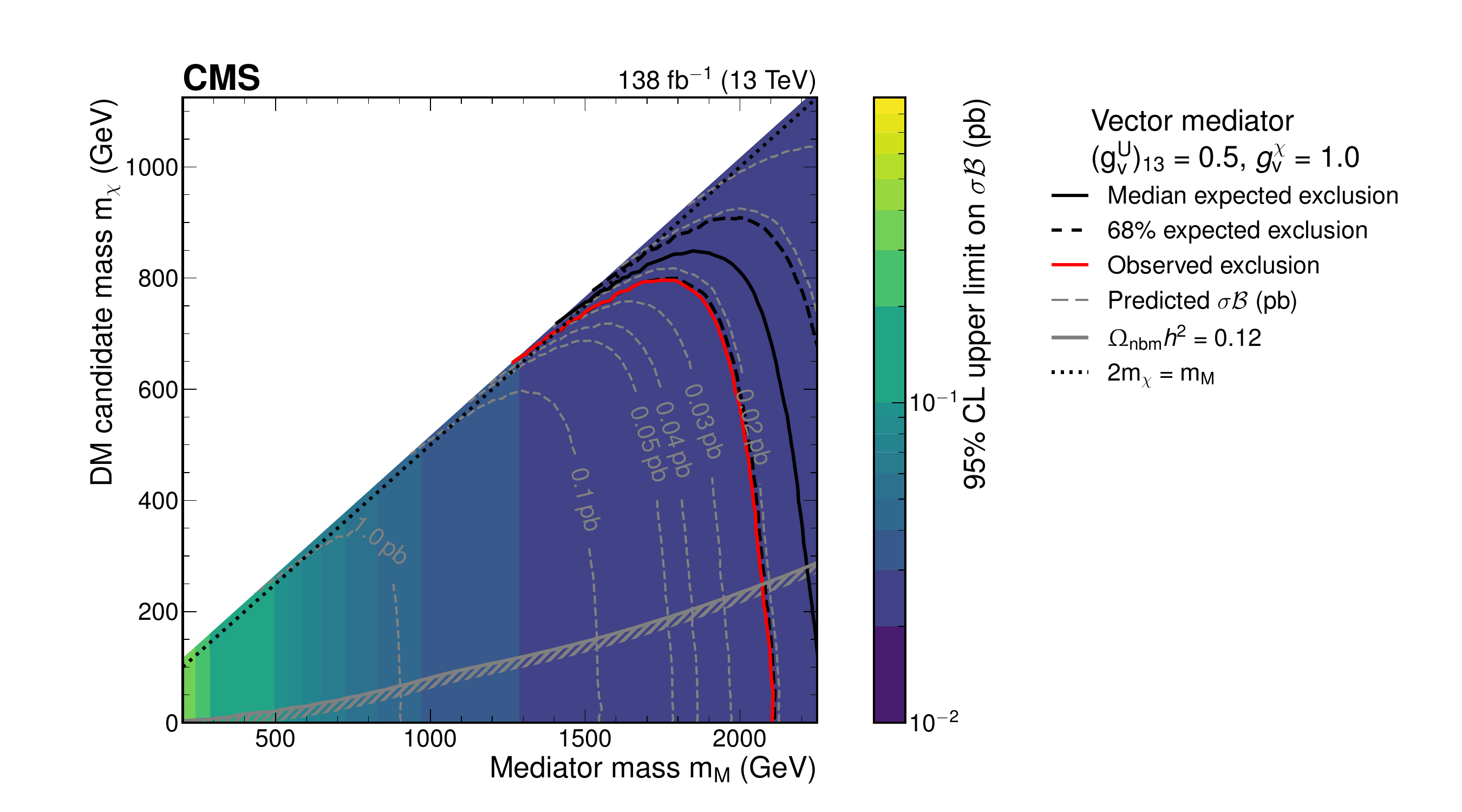}
  \\
  \includegraphics[width=0.95\textwidth]{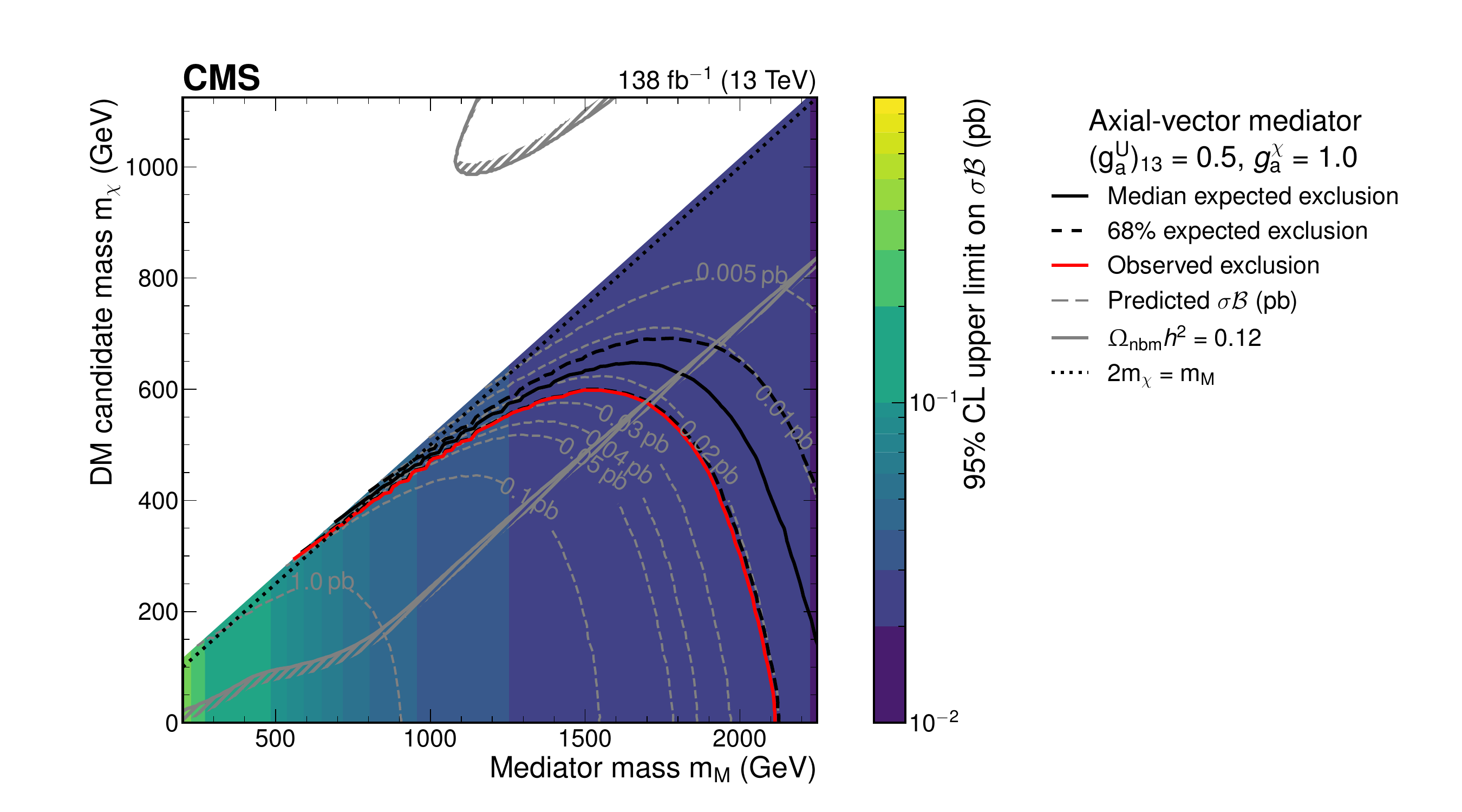}
  \caption{%
    Upper limits at 95\% \CL on \sigmaB of mono-top production presented in the two-dimensional plane spanned by the mediator and DM candidate masses for a mediator mass between 200 and 2250\GeV and a DM candidate mass between 1 and 1125\GeV only considering on-shell decays of the mediator to the DM candidates.
    The mediator has vector couplings to quarks and DM candidates in the upper plot and axial-vector couplings in the lower plot.
    The median expected exclusion range is indicated by a black solid line, demonstrating the search sensitivity of the analysis.
    The 68\% probability interval of the expected exclusion is shown in black dashed lines.
    Contours of theory predictions for constant values of \sigmaB are shown in gray dashed lines.
    The observed exclusion contour of mediator and DM candidate masses is represented by the red solid line.
    The exclusion contour obtained from measurements of the DM relic density \relicdensity by the Planck Collaboration is shown in the gray solid line.
  }
  \label{fig:hadResults:all:AltCoupling}
\end{figure}
\cleardoublepage \section{The CMS Collaboration \label{app:collab}}\begin{sloppypar}\hyphenpenalty=5000\widowpenalty=500\clubpenalty=5000
\cmsinstitute{Yerevan Physics Institute, Yerevan, Armenia}
{\tolerance=6000
V.~Chekhovsky, A.~Hayrapetyan, V.~Makarenko\cmsorcid{0000-0002-8406-8605}, A.~Tumasyan\cmsAuthorMark{1}\cmsorcid{0009-0000-0684-6742}
\par}
\cmsinstitute{Institut f\"{u}r Hochenergiephysik, Vienna, Austria}
{\tolerance=6000
W.~Adam\cmsorcid{0000-0001-9099-4341}, J.W.~Andrejkovic, L.~Benato\cmsorcid{0000-0001-5135-7489}, T.~Bergauer\cmsorcid{0000-0002-5786-0293}, S.~Chatterjee\cmsorcid{0000-0003-2660-0349}, K.~Damanakis\cmsorcid{0000-0001-5389-2872}, M.~Dragicevic\cmsorcid{0000-0003-1967-6783}, P.S.~Hussain\cmsorcid{0000-0002-4825-5278}, M.~Jeitler\cmsAuthorMark{2}\cmsorcid{0000-0002-5141-9560}, N.~Krammer\cmsorcid{0000-0002-0548-0985}, A.~Li\cmsorcid{0000-0002-4547-116X}, D.~Liko\cmsorcid{0000-0002-3380-473X}, I.~Mikulec\cmsorcid{0000-0003-0385-2746}, J.~Schieck\cmsAuthorMark{2}\cmsorcid{0000-0002-1058-8093}, R.~Sch\"{o}fbeck\cmsAuthorMark{2}\cmsorcid{0000-0002-2332-8784}, D.~Schwarz\cmsorcid{0000-0002-3821-7331}, M.~Sonawane\cmsorcid{0000-0003-0510-7010}, W.~Waltenberger\cmsorcid{0000-0002-6215-7228}, C.-E.~Wulz\cmsAuthorMark{2}\cmsorcid{0000-0001-9226-5812}
% ...
\par}
\vskip\cmsinstskip
\dag:~Deceased\\
$^{1}$Also at Yerevan State University, Yerevan, Armenia\\
$^{2}$Also at TU Wien, Vienna, Austria\\
% ...
\end{sloppypar}
%%% END EDITABLE REGION %%%
% skeleton_end
\end{document}